\documentclass[pra,aps,twocolumn,showpacs]{revtex4}
\usepackage{amsmath,graphicx,epsfig}

\def\prn#1{{\left(#1\right)}}

\def\ts#1{{_{\mbox{\scriptsize #1}}}}

\def\dbydt#1{{\frac{d #1}{dt}}}

\def\fig_width{3. in} 
\draft

\newlength{\defbaselineskip}
\newcommand{\setlinespacing}[1]%
           {\setlength{\baselineskip}{#1 \defbaselineskip}}

\begin{document}

\title{Electric-field-induced change of alkali-metal vapor density in paraffin-coated cells} 

\author{D.\ F.\ Jackson Kimball}
\email{derek.jacksonkimball@csueastbay.edu}
\author{Khoa Nguyen}
\altaffiliation{Present address: Department of Physics and Astronomy, San Jose State University, San Jose, California 95192-0106}
\affiliation{Department of Physics, California State University --
East Bay, Hayward, California 94542-3084}

\author{K. Ravi}
\author{Arijit Sharma}
\author{Vaibhav S. Prabhudesai}
\altaffiliation{Present address: Department of Physics and Complex Systems, Weizmann Institute of Science, Rehovot, Israel 76100.}
\author{S. A. Rangwala}
\affiliation{Raman Research Institute,
Sadashivanagar,
Bangalore 560080 India}

\author{V.\ V.\ Yashchuk}
\affiliation{Advanced Light Source Division, Lawrence Berkeley
National Laboratory, Berkeley, California 94720}

\author{M. V. Balabas}
\affiliation{S. I. Vavilov State Optical Institute, St. Petersburg, 199034 Russia}

\author{D. Budker}
\affiliation{Department of Physics, University of
California at Berkeley, Berkeley, California 94720-7300} \affiliation{Nuclear Science
Division, Lawrence Berkeley National Laboratory, Berkeley, California 94720}

\date{\today}



\begin{abstract}

Alkali vapor cells with antirelaxation coating (especially paraffin-coated cells) have been a central tool in optical pumping and atomic spectroscopy experiments for 50 years.  We have discovered a dramatic change of the alkali vapor density in a paraffin-coated cell upon application of an electric field to the cell.  A systematic experimental characterization of the phenomenon is carried out for electric fields ranging in strength from $0-8~{\rm kV/cm}$ for paraffin-coated cells containing rubidium and cells containing cesium. The typical response of the vapor density to a rapid (duration $\lesssim 100~{\rm ms}$) change in electric field of sufficient magnitude includes (a) a rapid (duration of $\lesssim 100~\rm{ms}$) and significant increase in alkali vapor density followed by (b) a less rapid (duration of $\sim 1~\rm{s}$) and significant decrease in vapor density (below the equilibrium vapor density), and then (c) a slow (duration of $\sim 100~\rm{s}$) recovery of the vapor density to its equilibrium value.  Measurements conducted after the alkali vapor density has returned to its equilibrium value indicate minimal change (at the level of $\lesssim$10\%) in the relaxation rate of atomic polarization.  Experiments suggest that the phenomenon is related to an electric-field-induced modification of the paraffin coating.


\end{abstract}




\maketitle

\section{Introduction}

Considerable efforts have been undertaken over many years to develop experimental techniques to reduce the relaxation rate of coherences between atomic states.  Long-lived atomic coherences enable precise measurements of atomic properties, enhance nonlinear optical effects, and are the key to ultra-precise atomic clocks and magnetometers.  One technique that has been particularly effective in preserving Zeeman and hyperfine coherences in alkali atoms has been to contain alkali vapors in evacuated (buffer-gas free) glass cells whose inner surfaces are coated with paraffin ($\rm{C_nH_{2n+2}}$, where in our work $\rm{n} \sim 40-60$).  Coating the walls of a vapor cell with paraffin can reduce the spin relaxation rate due to wall collisions by four orders of magnitude \cite{Rob58,Bou66,Ale92,Ale96}. The narrow resonances ($\sim 1$~Hz) that result have been applied in sensitive magnetometry \cite{Dup69,Ale92,Ale96,Bud98,Bud00sens,Bud02_FM_NMOR,Ale04,Aco06}, the study of electromagnetically induced transparency and light propagation dynamics \cite{Bud99,Wan00}, the study of high-order Zeeman coherences \cite{Yas03,Pus06,Aco08}, and tests of fundamental symmetries \cite{Ens62,Ens67,Eks71,Kim01,Kim07}.

One important class of experiments to which paraffin-coated alkali vapor cells have not been widely applied are those involving electric fields, in particular searches for the parity and time-reversal violating permanent electric dipole moment (EDM) of the electron $d_e$.  There were, in fact, a series of experimental searches for the electron EDM \cite{Ens62,Ens67,Eks71} carried out at the University of Washington using alkali atoms contained in paraffin-coated cells during the 1960's and early 1970's that set a limit of $|d_e| < 1.6 \times 10^{-23}~{\rm e \cdot cm}$ \cite{Eks71}.  These experiments were limited by a systematic effect involving an alteration of the cell properties due to application of the electric fields.  Subsequent electron EDM searches using other methods have far surpassed this level of sensitivity, with the latest effort employing a thallium atomic beam resulting in a limit of $|d_e| < 1.6 \times 10^{-27}~{\rm e \cdot cm}$ \cite{Reg02}.  Nonetheless, recent developments indicated that an electron EDM search employing paraffin-coated cells may, after all, be able to offer some hope of competing with the latest generation of EDM experiments. Combining state-of-the-art paraffin-coated cell technology \cite{Ale02} with the use of nonlinear optical rotation for detection of spin precession \cite{Bud98,Bud00sens,Bud02_FM_NMOR}, and assuming that an electric field of $\sim 10~{\rm kV/cm}$ can be applied to cesium atoms contained in a paraffin-coated cell, one obtains an estimate for the shot-noise-limited statistical sensitivity of such an experiment to the electron EDM of $\delta|d_e| \sim 10^{-26}~{\rm e \cdot cm}/\sqrt{\rm Hz}$ \cite{Yas99} (taking into account the factor of 120 enhancement of the Cs EDM relative to the electron EDM \cite{San66,Joh86,Das88,Har90}).  Motivated by this possibility, we initiated a study of the application of electric fields to paraffin-coated cells.

Unexpectedly, we discovered a dramatic effect of the electric field on the alkali vapor density in paraffin-coated cells \cite{Kim01}.  For sufficiently large electric fields, when the electric field magnitude was altered or if its polarity was reversed, we observed a significant and rapid (time scale $\sim 100~{\rm ms}$) increase in the alkali vapor density, then a significant and rapid (time scale $\sim 1~{\rm s}$) decrease in the vapor density (well below the initial equilibrium density), followed by a relatively slow (time scale on the order of 100~s) return to roughly the initial vapor density.  Experiments performed with uncoated vapor cells showed no such effects, so the behavior is attributed to the paraffin coating.

In this paper we report on a series of experiments performed to characterize this phenomenon and study its dependence on electric field magnitude, cell temperature, and cell history.  This paper combines complementary sets of data taken using different experimental setups, procedures, geometries, cells, and atoms at three different institutions: California State University - East Bay (CSU-EB), the Raman Research Institute (RRI), and the University of California at Berkeley (UCB). We will begin by describing the experimental setups and the data acquired at each of the institutions, and then discuss possible interpretations of the results and physical mechanisms for the phenomenon.  We demonstrate that the most important property of the paraffin-coated cell, its ability to significantly reduce the relaxation rate of ground-state atomic coherences, is largely unaffected by the application of electric fields.  The phenomenon can be described using a rate equation model introduced previously \cite{Ale02} to understand the phenomenon of light-induced atomic desorption (LIAD) from paraffin.

Although electric-field-induced density changes have been observed in several different paraffin-coated cells and all paraffin-coated cells tested, it is possible that there may be cell-to-cell variation.  Note that it is possible that details of the effect depend on the cell preparation procedure, which for the cells employed in the described experiments is discussed in Ref.~\cite{Ale02}.

\section{Experiments with Cesium}\label{Sec:Cesium}

\subsection{Experimental Setup}

\begin{figure}
\center
\includegraphics[width=3.35 in]{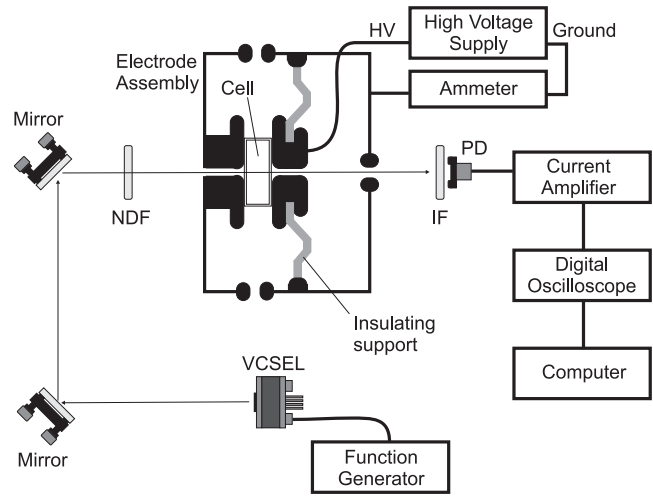}
\caption{Schematic of setup for experiments with cesium atoms contained in paraffin-coated cells at California State University - East Bay (CSU-EB). HV - high voltage output of power supply, NDF - neutral density filter, IF - interference filter, PD - photodiode. }\label{Fig:ExpSetupCSUEB}
\end{figure}

The setup for the experiments carried out at CSU-EB is shown in Fig.~\ref{Fig:ExpSetupCSUEB}. The paraffin-coated cells used in these experiments were 3-cm tall, 6-cm diameter cylinders with a single stem containing a droplet of Cs metal.  The cells were evacuated to $\sim 10^{-5}~{\rm Torr}$ at the time of manufacturing and do not contain any buffer gas.  The Cs vapor density was monitored by measuring absorption spectra of light resonant with the D2 (852 nm) line.  The light was generated by a vertical-cavity surface emitting laser (VCSEL) and the light frequency was scanned over 10 GHz to cover the entire hyperfine structure by modulating the current of the VCSEL with a triangular waveform.  The intensity of the light was maintained below $\approx 10~\rm{\mu W/cm^2}$ with use of neutral density filters in order to minimize distortion of the lineshape due to optical pumping (optical pumping effects can be important even at low light powers in paraffin-coated cells because of the long lifetime of ground-state coherences). Transmitted light power was detected by a photodiode fitted with an interference filter (peak transmission at $850~{\rm nm}$ with a $10~{\rm nm}$ bandpass) to reduce background noise from stray light.  The photodiode was connected to a current pre-amplifier and the output was recorded by a digital oscilloscope. Every five seconds an oscilloscope trace was transferred to a computer where the data were fit in real time to an absorption spectrum based on the Voigt profile in order to extract the Cs vapor density.

\begin{figure}
\center
\includegraphics[width=3.35 in]{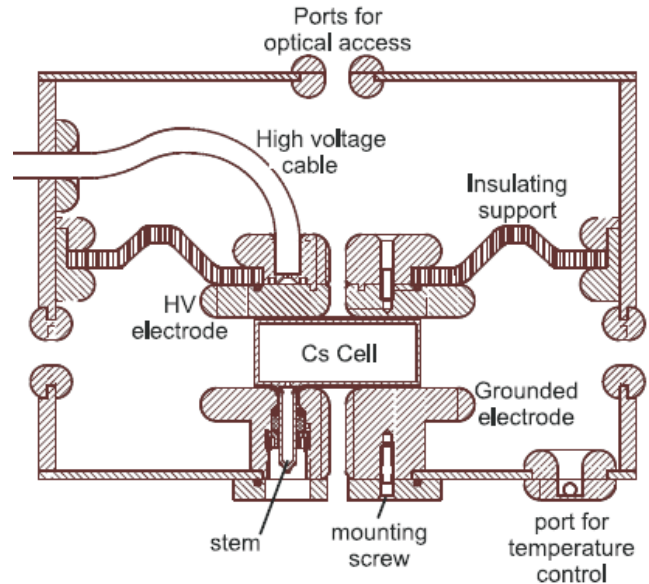}
\caption{Design of high voltage electrode assembly for the experiment.  Cylindrical box is made of copper, rounded edges are used on surfaces near the high voltage electrode to minimize possibility of discharge.  Insulating support for the high voltage plate is manufactured from G10 plastic.}\label{Fig:HV electrodes}
\end{figure}

The paraffin-coated cell was mounted inside the copper electrode assembly shown in Fig.~\ref{Fig:HV electrodes}.  The power supply can reverse polarity with a roughly 1-second time constant.  Voltages of up to 15~kV were applied to the high voltage electrode without any evidence of breakdown.  Leakage current between the plates was monitored with a precision ammeter connected in series across the path to the ground of the power supply, and within the sensitivity of our instrument (10 pA) we found no evidence of leakage current through the cell.  Experiments reported previously \cite{Bud02_Bennett_structures} used electro-optic effects to verify that the applied electric field was present inside the volume of the cell with the correct magnitude (measurement accuracy was within a factor of two).  The temperature of the cell was controlled by blowing heated or cooled air into the electrode assembly through the port for temperature control (see Fig.~\ref{Fig:HV electrodes}).  In some experiments, we also independently controlled the temperature of the cell stem by blowing heated or cooled air directly onto the stem while the rest of the cell remained at room temperature ($\approx 20^\circ$C).  The external flat faces of the cell were coated with silver paint and a thin layer of conductive silicone (Cho-Bond 1038) was affixed to the surfaces of the electrode plates to ensure good electrical contact between the plates and the outer surface of the cell.  Two different cells of nearly identical geometry manufactured at the same time were employed in these experiments, with nearly identical results.  During the experiments, the laboratory magnetic field was uncontrolled, and has since been measured to be of typical magnitude $0.4~{\rm G}$.  During the experiments, the orientation of the setup was changed from time-to-time relative to the laboratory magnetic field with no noticeable change in the experimental results.

Note that the dc Stark shift of the optical transition frequency does not affect the determination of vapor density in the present scheme since we fit the entire Doppler-broadened absorption spectrum.  Furthermore, even at the highest electric field magnitudes, the dc Stark shift of the optical transition frequency ($\sim 10~{\rm MHz}$, see Refs.~\cite{Hun88,Tan88}) is a small fraction of the Doppler width ($\sim \rm 400~{\rm MHz}$).

\subsection{Observations}

\begin{figure}
\center
\includegraphics[width=3.35 in]{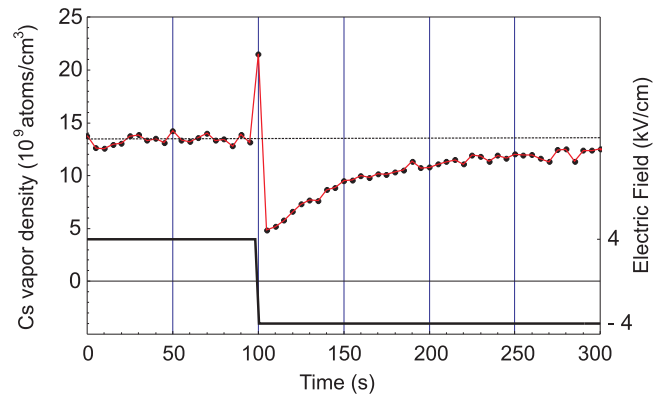}
\caption{Cesium density as a function of time (upper plot, data points connected with line) with electric field of 4~kV/cm magnitude applied (lower plot).  The polarity of the electric field is switched at $t=100~{\rm s}$.  Dashed line indicates average equilibrium density $n_0$ prior to field reversal. Temperature of the cell and stem are $\approx 20^\circ$C.  Density data acquired every 5 seconds, extracted from a fit to the Cs D2 absorption spectrum. Data acquisition is too slow to show the detailed dynamics of the rapid increase and decrease of vapor density as the polarity is switched, although subsequent experiments at RRI using a different method of data acquisition are able to investigate these dynamics. Fluctuations in the value of $n$ are due to frequency jitter of the VCSEL during scans through the absorption spectrum, which degrades the quality of the fits.}\label{Fig:Efield_reversal_4kV_Zoom}
\end{figure}

The basic character of the Cs vapor density variation with application of time-dependent electric fields to the parrafin-coated cell is shown in Fig.~\ref{Fig:Efield_reversal_4kV_Zoom}.  When the electric field polarity is switched, at first there is a rapid increase in Cs vapor density with a time scale of $\sim$~1~s, which we term the ``burst'' of Cs vapor density.  Although the speed of the data acquisition in the experimental setup at CSU-EB was too slow to reliably capture the details of the Cs vapor burst, the experiments conducted at RRI with Rb are able to probe these relatively fast dynamics, so we defer detailed discussion of the burst until later.  We observed burst amplitudes of more than twice the initial equilibrium vapor density at the highest electric fields used ($5~{\rm kV/cm}$).

After the short-lived burst of Cs atoms, the vapor density rapidly declines far below the initial equilibrium vapor density (we term this the ``drop'' in Cs density), and then recovers relatively slowly (time scale $\sim 100~{\rm s}$).  Note that for the data shown in Fig.~\ref{Fig:Efield_reversal_4kV_Zoom}, the density does not fully recover to the equilibrium value $n_0$ (dashed line) prior to the switching of the field polarity.  This general decline of the equilibrium vapor density was consistently observed as the field polarity was repeatedly switched.  Full recovery could take up to over an hour, as will be discussed in some detail below.  Due to the limitations of our present data acquisition system, the experiments at CSU-EB focused primarily on the character of the drop in Cs vapor density.

It is also of interest to note that, as has been commonly observed in paraffin-coated alkali vapor cells (see, for example, Refs.~\cite{Bou66,Lib86,Bal93,Bal95,Ale02}), the equilibrium Cs vapor density prior to application of the electric field in both cells at room temperature (20$^\circ$C) was $\approx$ 50\% lower than the saturated vapor density for Cs of $\approx 3.0 \times 10^{10}~{\rm cm^{-3}}$ \cite{CRC}. This suppression of the alkali vapor densities has generally been attributed to the continuous adsorption of atoms into the paraffin coating \cite{Bou66,Lib86,Bal93,Bal95}. There is direct evidence that alkali atoms become trapped in the paraffin coating from experiments in which light is used to desorb alkali atoms from the coating (a phenomenon known as light-induced atomic desorption [LIAD], see Refs.~\cite{Ale02,Gra05,Goz08,Kar08} for a discussion of LIAD from paraffin-coated surfaces).

\begin{figure}
\center
\includegraphics[width=3.35 in]{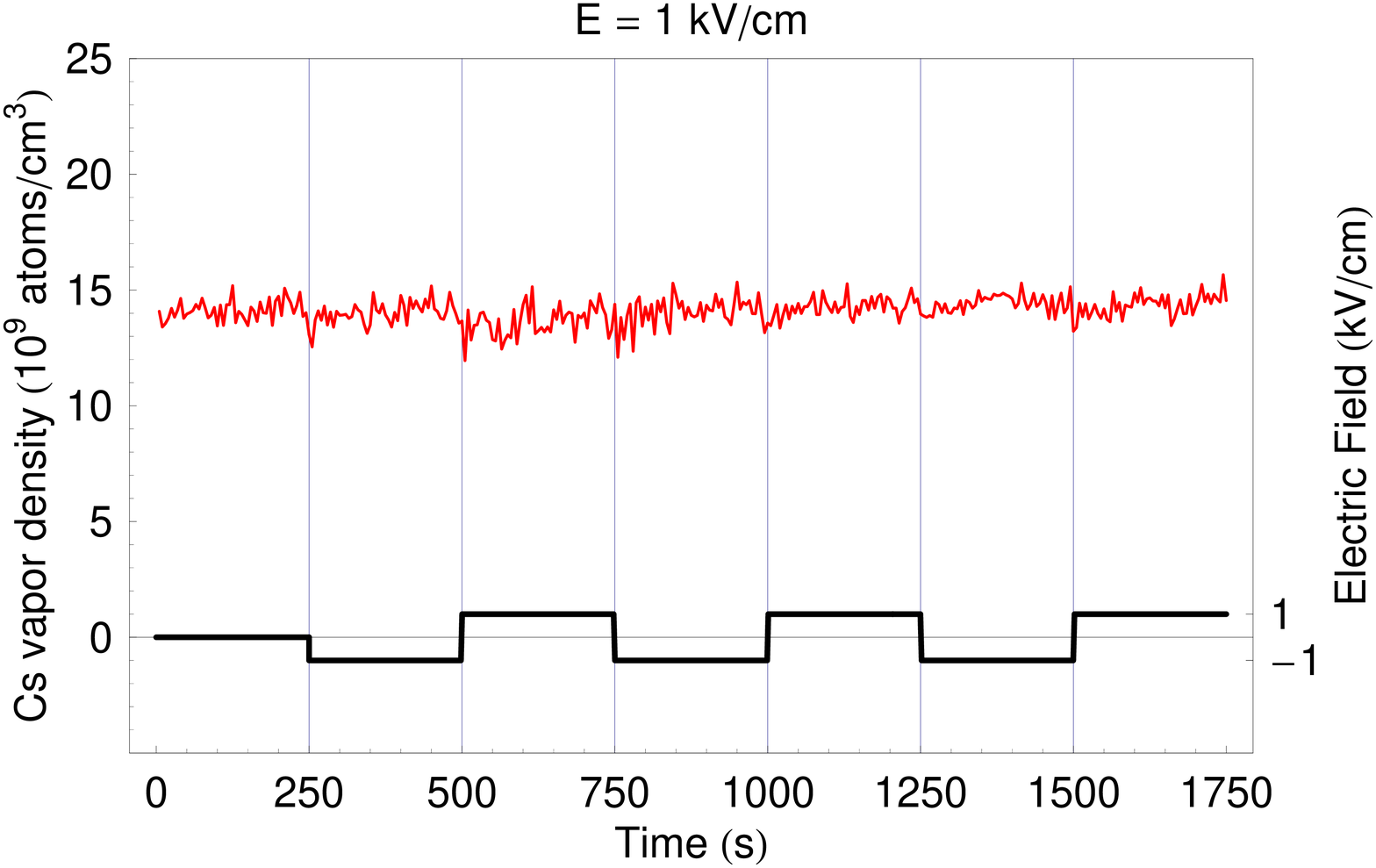}
\includegraphics[width=3.35 in]{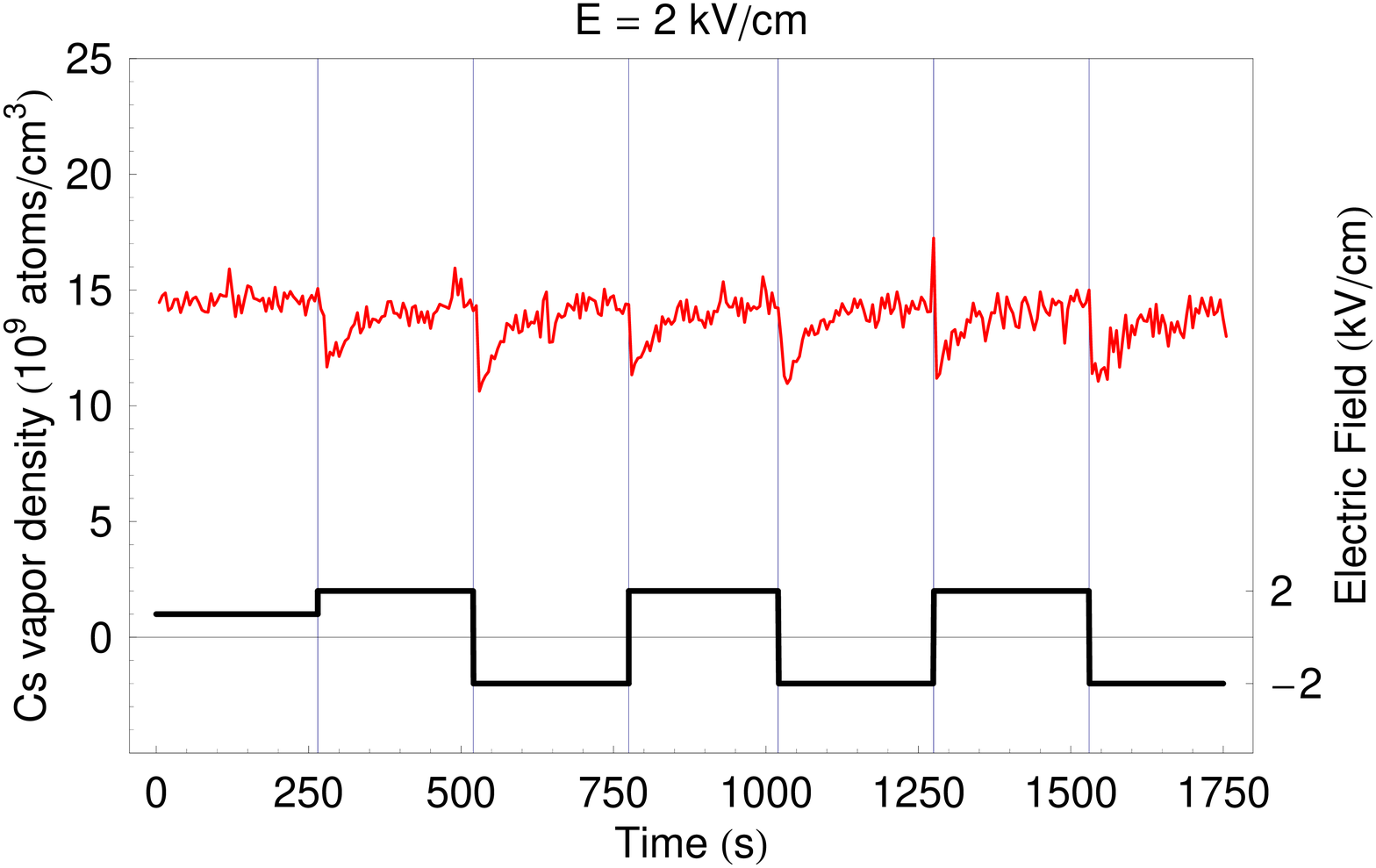}
\includegraphics[width=3.35 in]{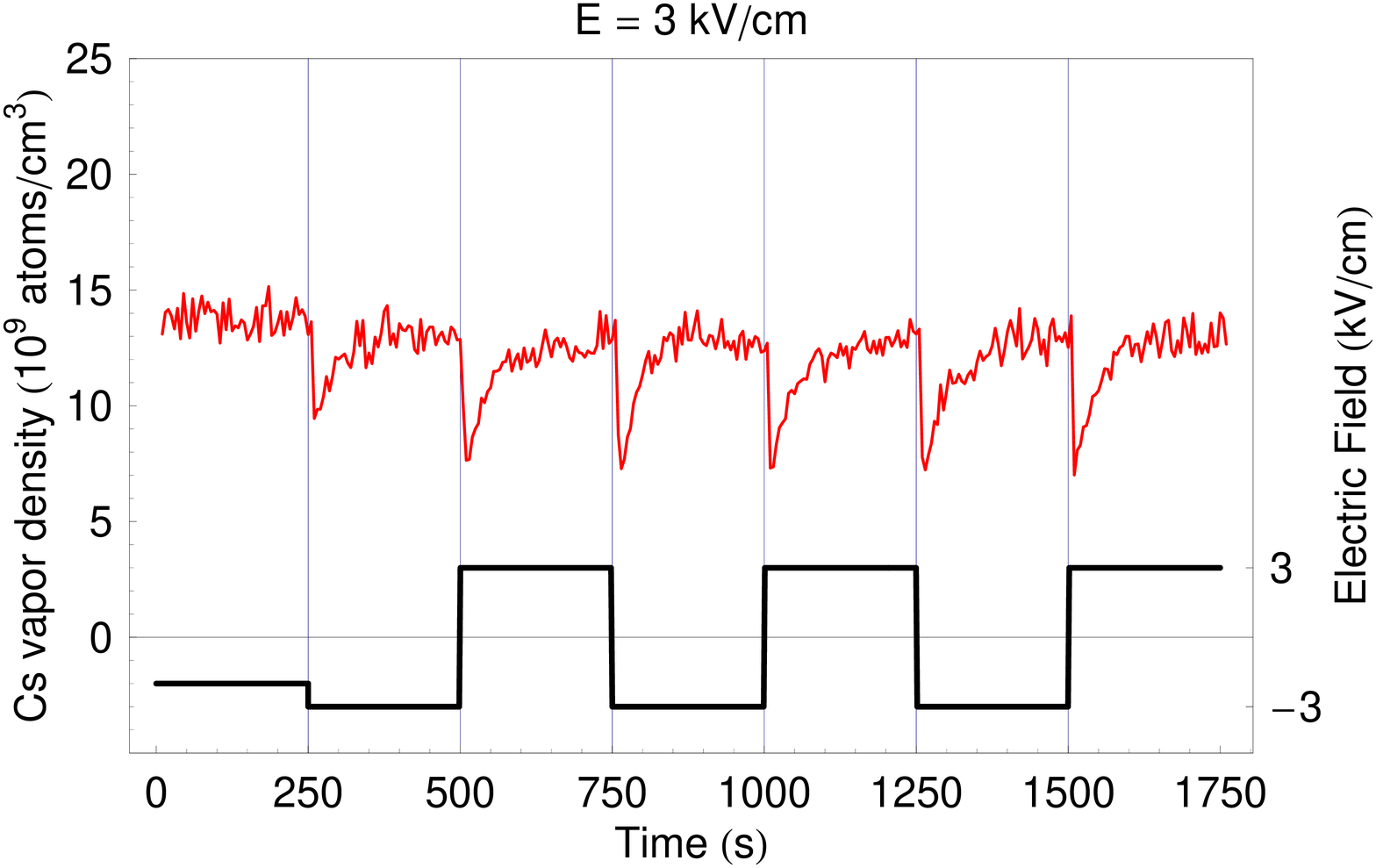}
\includegraphics[width=3.35 in]{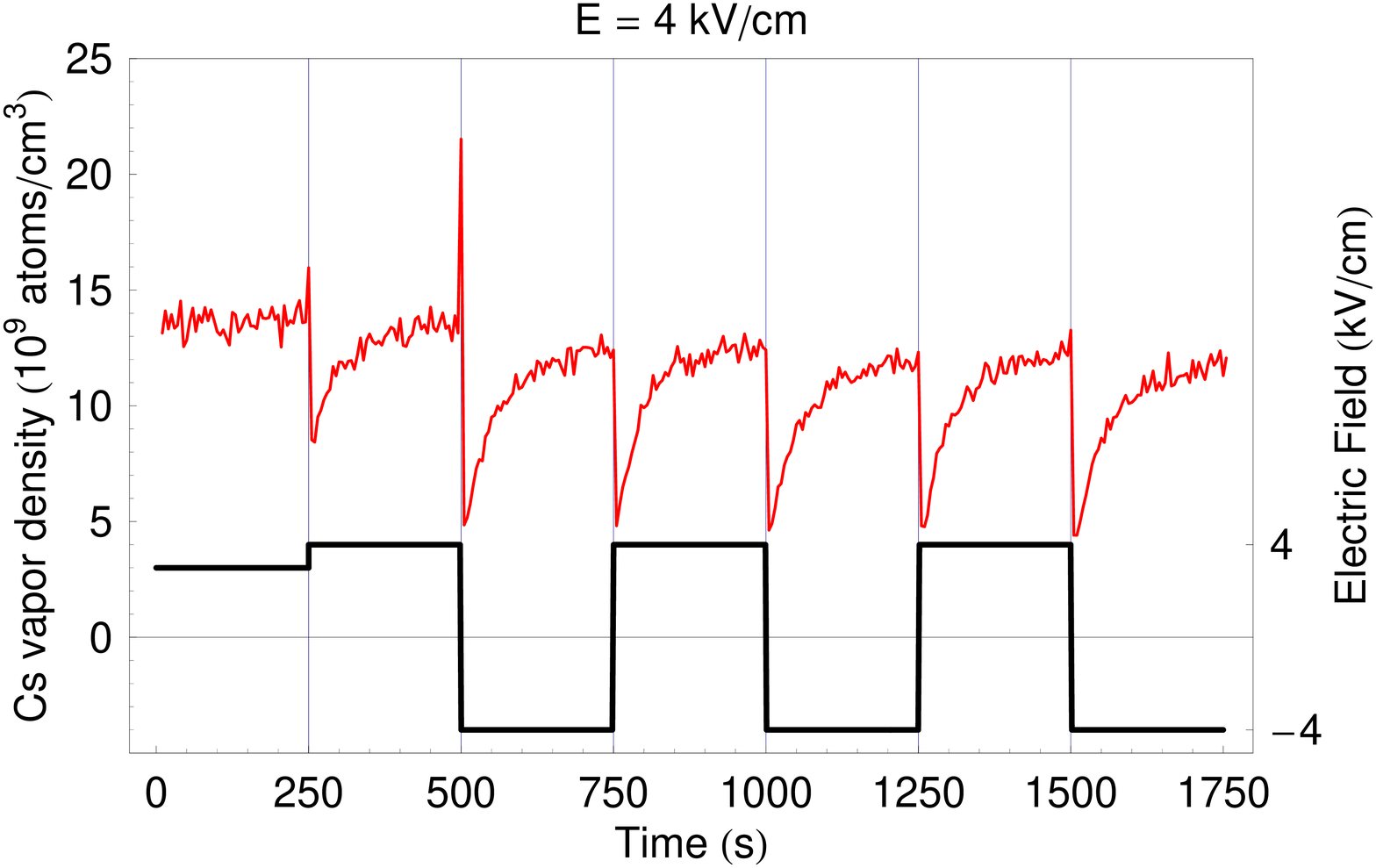}
\caption{Cesium density as a function of time (upper plots in each graph) with electric field of 1, 2, 3, and 4~kV/cm magnitude applied, polarity switched every 250~s (lower plots in each graph). The cell temperature was $20^\circ$C.}\label{Fig:Density_Efield_TimePlots_20C}
\end{figure}

Figure~\ref{Fig:Density_Efield_TimePlots_20C} shows the time dependence of the Cs vapor density in the paraffin-coated cell as the field polarity was repeatedly switched for different magnitudes of the electric field.  For this set of data, a sequence of polarity reversals was first performed at 1~kV/cm, then the field magnitude was subsequently increased in steps of 1~kV/cm and the procedure repeated.  The plots shown in Fig.~\ref{Fig:Density_Efield_TimePlots_20C} are in sequential order from top to bottom, so that the electric field in each plot begins at the value where the field ends in the plot above.  In addition to the systematic drop in density upon polarity reversals for fields above $2~{\rm kV/cm}$, one can also observe similar, but slightly smaller, changes in the Cs vapor density at the start of each sequence when the electric field magnitude was changed.  A general decline in the equilibrium Cs vapor density $n_0$  prior to a particular polarity reversal can be seen as the experimental run continues (especially pronounced for the data at $E = \pm 4~{\rm kV/cm}$), indicating that the cell did not fully recover to its initial state between polarity reversals.  Nonetheless, the properties of the drop can be characterized in terms of the change in vapor density $\Delta n$ from its value $n_0$ before a polarity reversal to after the reversal, as well as the fractional change in density $\Delta n/n_0$.  Although data acquisition is too slow to reliably capture the brief but significant increase (burst) in atomic density when the field is switched, however the burst is captured at $t = 1275~{\rm s}$ for the 2~kV/cm data and at $t=500~{\rm s}$ for the 4~kV/cm data.

\begin{figure}
\center
\includegraphics[width=3.35 in]{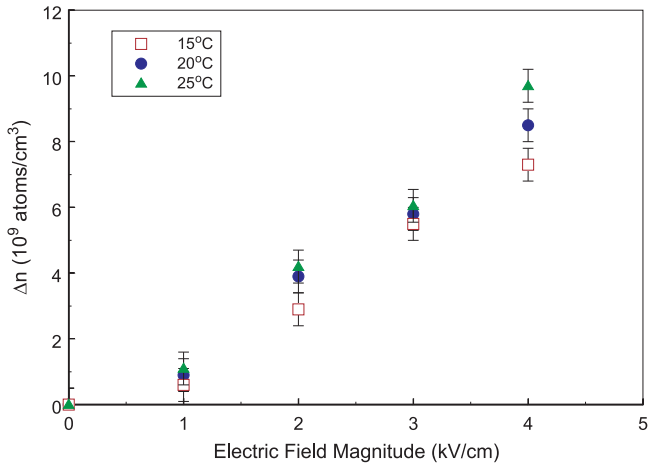}
\caption{Average maximum decrease in Cs vapor density ($\Delta n$) when field polarity is switched as a function of electric field magnitude for different cell temperatures.  Electric field magnitude starts at zero and is increased sequentially in steps of 1~kV/cm. There was a 30 minute delay between changes in field magnitude, and data at different temperatures were taken on different days.}\label{Fig:DensityChange_vs_Efield}
\end{figure}

\begin{figure}
\center
\includegraphics[width=3.35 in]{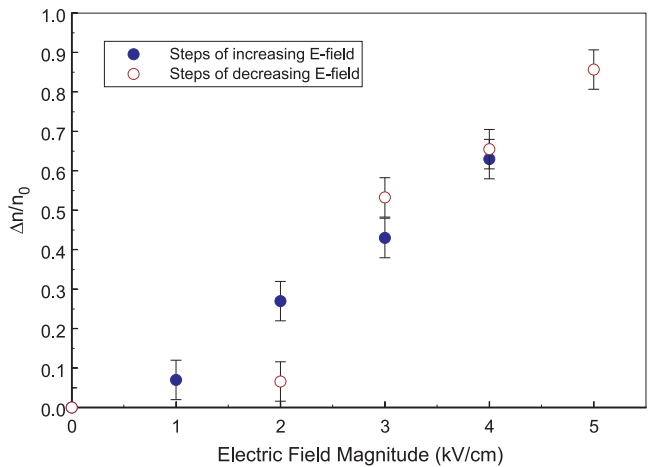}
\caption{Average maximum fractional decrease in Cs vapor density ($\Delta n/n_0$) when field polarity is switched as a function of electric field magnitude for the case where the electric field magnitude is successively increased (filled circles) and the case where the electric field magnitude is successively decreased (hollow circles).  Data taken at $T=20^\circ$C, with a 30 minute delay between changes in field magnitude.}
\label{Fig:FractionalDensityChange_vs_Efield_rev}
\end{figure}

One readily apparent feature of the phenomenon revealed by these measurements is the increase in the amplitude of the drop $\Delta n$ as the magnitude of the field is increased.  In particular it is of interest to note that at 1~kV/cm there is minimal change in Cs vapor density, whereas a significant drop is observed for polarity reversals at 2~kV/cm, possibly indicating a threshold for the effect, or at least a nonlinear increase in the effect for electric field magnitudes between 0 -- 2~${\rm kV/cm}$.  Figure~\ref{Fig:DensityChange_vs_Efield} shows the drop in density $\Delta n$ due to polarity reversals as a function of electric field magnitude for different cell temperatures.  The values for $\Delta n$ shown in Fig.~\ref{Fig:DensityChange_vs_Efield} are calculated from the difference between the average of 30~s of data prior to a polarity reversal and the average of 5~s of data (two points) after a reversal, and each point in Fig.~\ref{Fig:DensityChange_vs_Efield} represents the average of five polarity reversals.  For each set of data, the electric field is sequentially increased from low to high value during the experiment.  It is of interest to note that although the initial equilibrium Cs vapor density prior to application of the electric field changes by a factor of three between the lowest temperature and the highest temperature studied, $\Delta n$ is nearly the same for all the studied temperatures.  Also, it is important to note that at the highest electric field magnitudes, $\Delta n$ is approaching the initial density in the cell ($\Delta n/n_0$ is approaching 1, a 100\% drop in density), so the linear trend observed in the data cannot continue for significantly higher electric field magnitudes.

We also performed experiments in which the stem of the cell was locally heated (up to $30^\circ$C) while the rest of the cell was maintained at a temperature of $20^\circ$C, which gave nearly the same results for $\Delta n$ as experiments in which the entire cell was heated.  Heating the stem of the cell produced approximately the same vapor density in the cell as heating the whole cell, a consequence of the equilibrium reached between the alkali atoms from the stem and relatively slow and continuous adsorption of alkali atoms into the paraffin coating, observed in several previous studies \cite{Ale02,Lib86,Bal93,Bal95}.  (It is generally inadvisable to overheat the stem relative to the cell volume for extended periods of time, as this has been observed to degrade the relaxation properties of the paraffin coating by creating alkali metal deposits on the inner surface of the cell that act as relaxation sites \cite{Bou66}.  In our case, the stem was heated by no more than $10^\circ$C for only a few hours and there was no evidence of coating degradation after the experiment.)  The invariance of $\Delta n$ with respect to the alkali vapor density and the temperature of the cell suggest that the physical origin of the phenomenon is related to some property of the paraffin coating that is independent of both experimental parameters over the studied range.

In addition, we analyzed the time constant for recovery from the drop of Cs vapor density (taken to be the time for density to reach $n_0 - \Delta n/2$ after a polarity reversal) as a function of all the varied experimental parameters: electric field magnitude, cell temperature (Cs vapor density), etc.  No statistically significant dependence of the time constant on any of the parameters was found.  (Attempts to fit the time-dependent vapor density revealed that there were multiple exponentials involved in the recovery, and without a reliable theory of the effect it was difficult to extract reasonable, linearly independent fit parameters from the data.)

\begin{figure}
\center
\includegraphics[width=3.25 in,height=1.6 in]{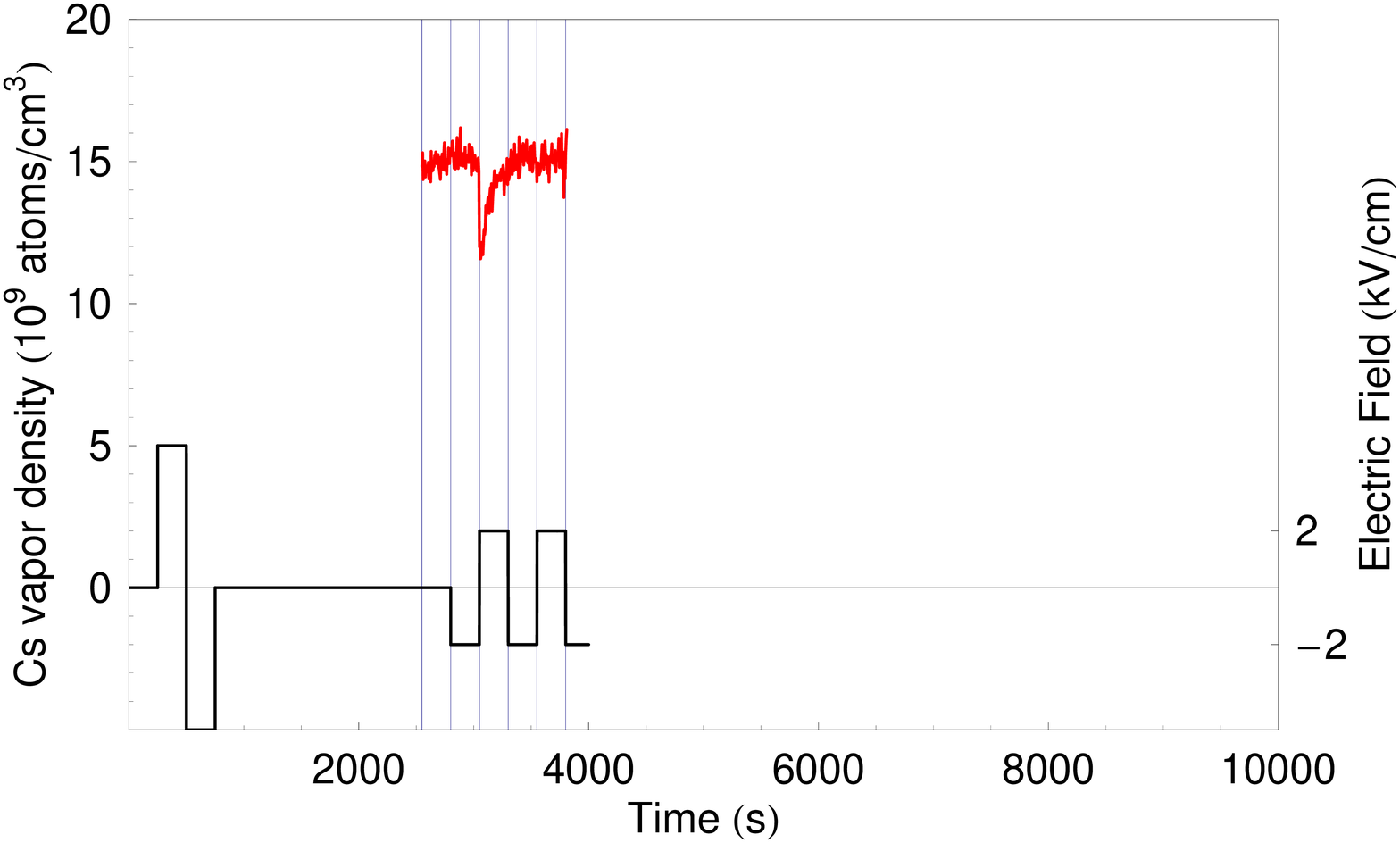}
\includegraphics[width=3.25 in,height=1.6 in]{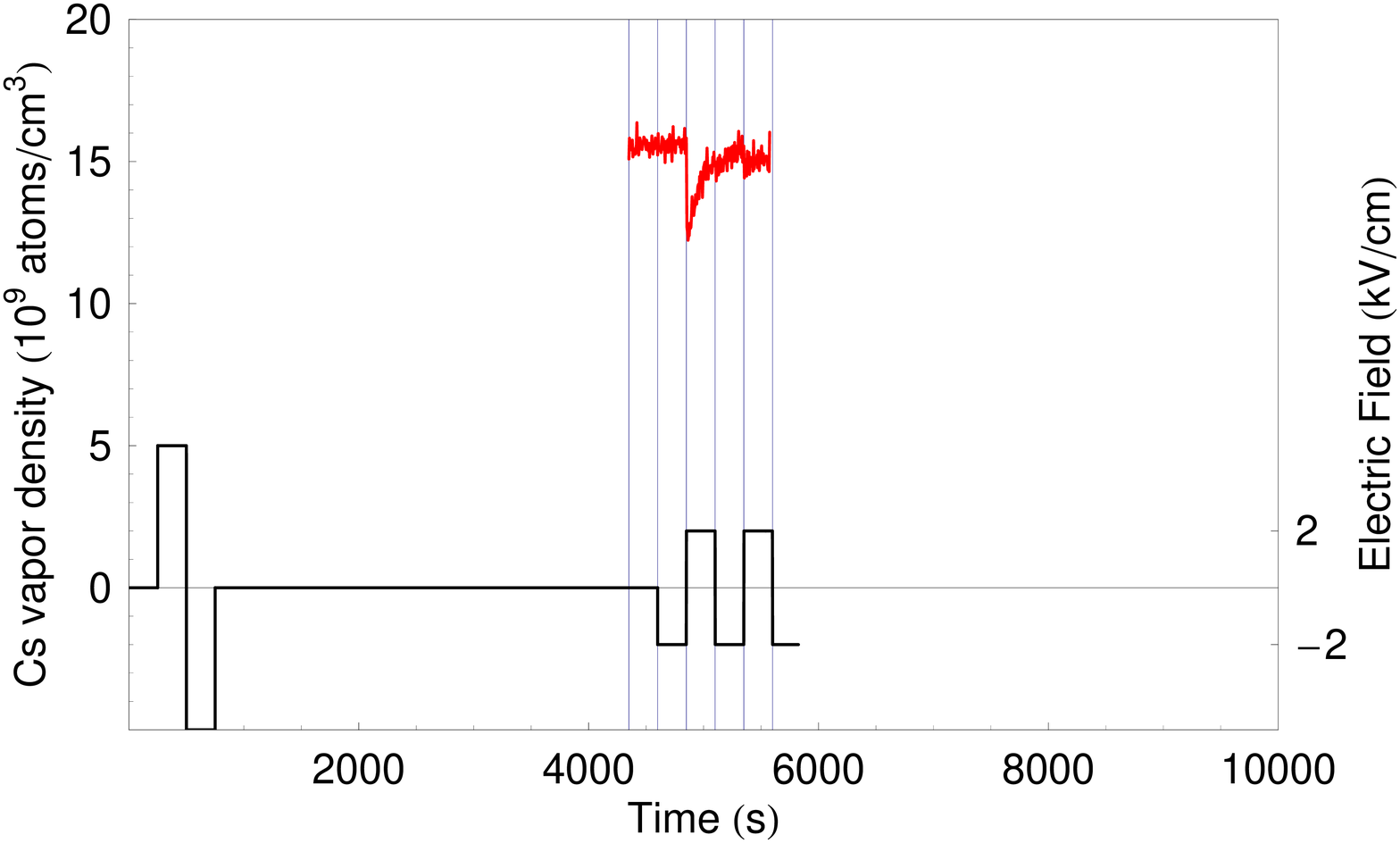}
\includegraphics[width=3.25 in,height=1.6 in]{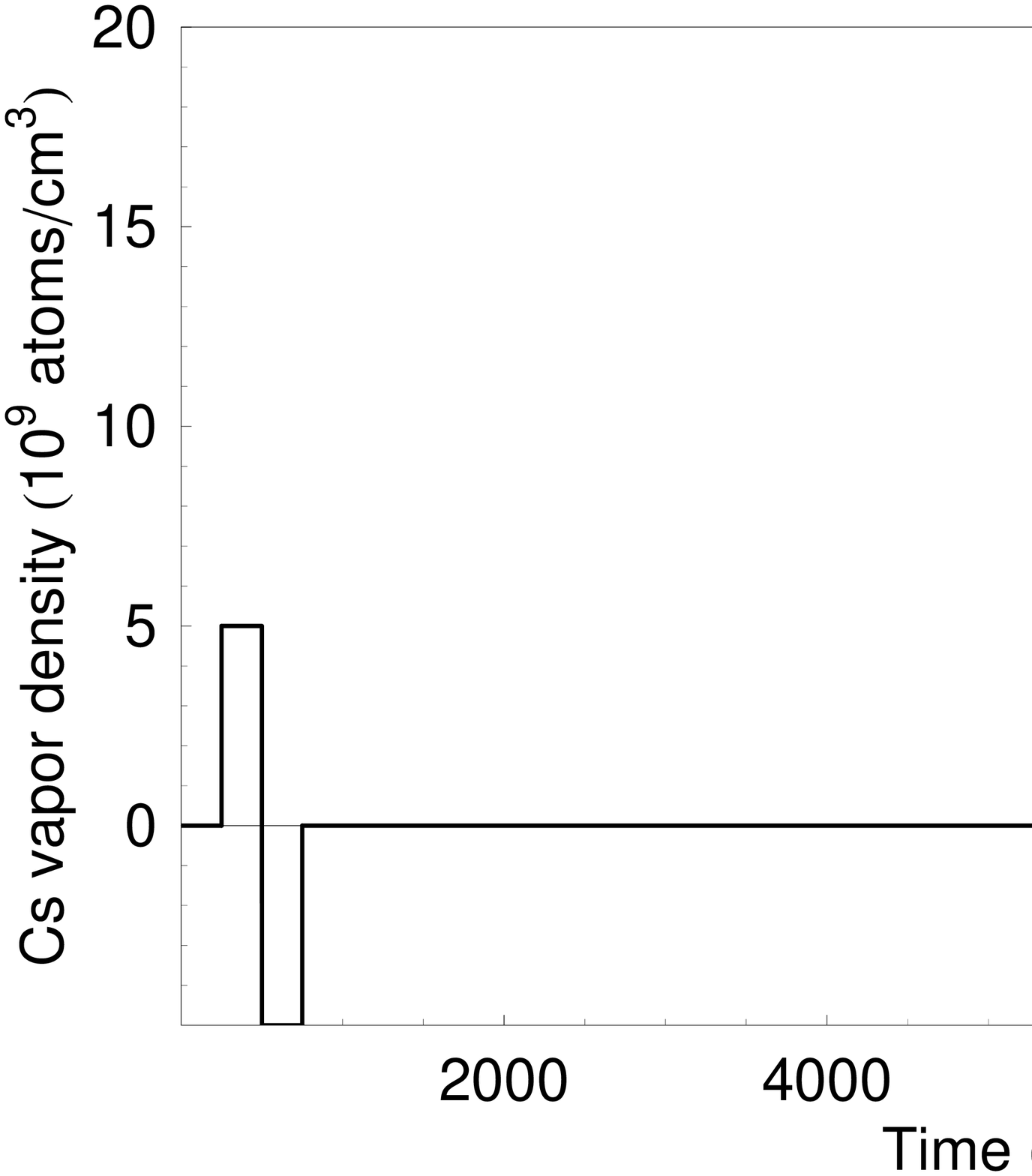}
\includegraphics[width=3.25 in,height=1.6 in]{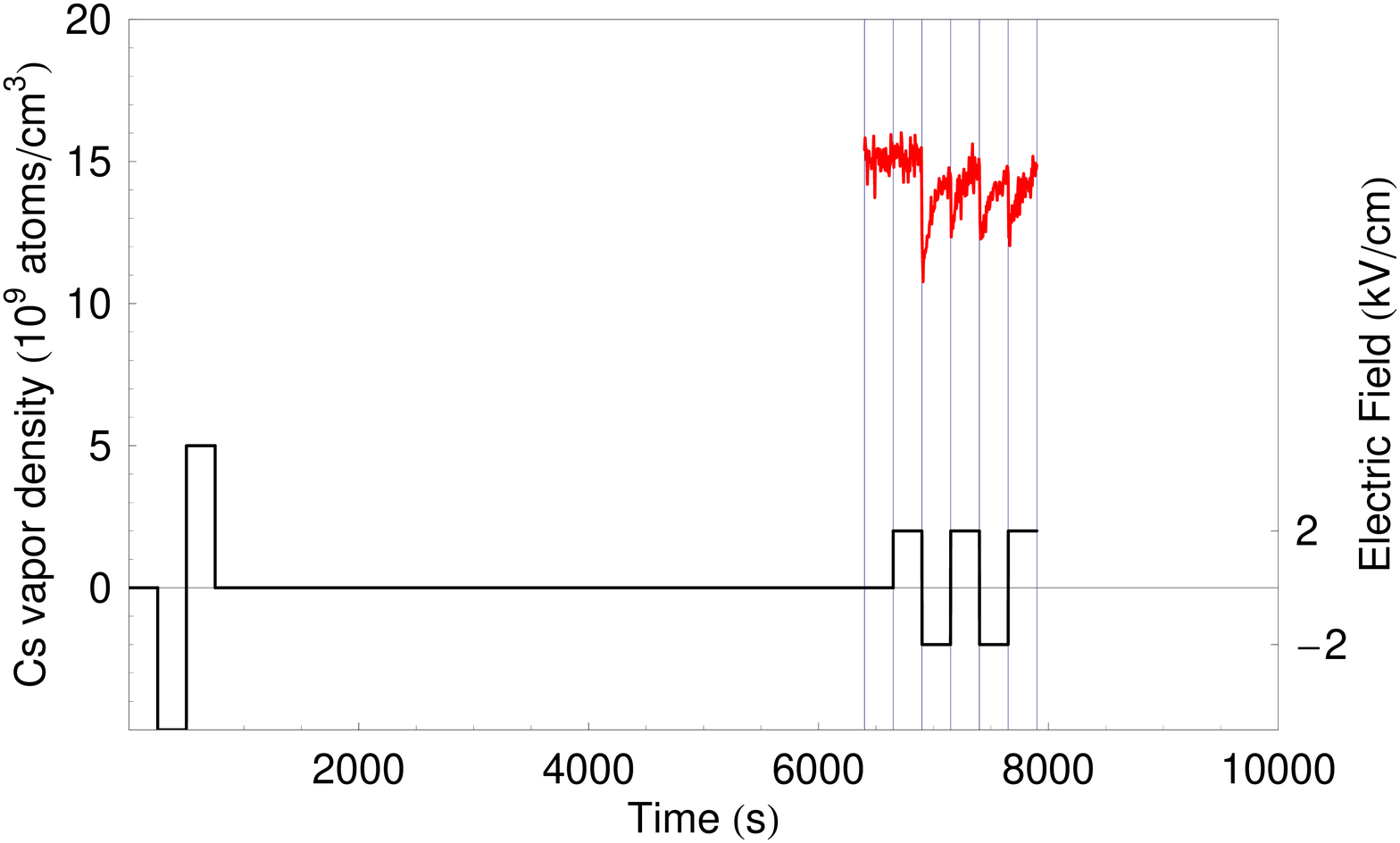}
\includegraphics[width=3.25 in,height=1.6 in]{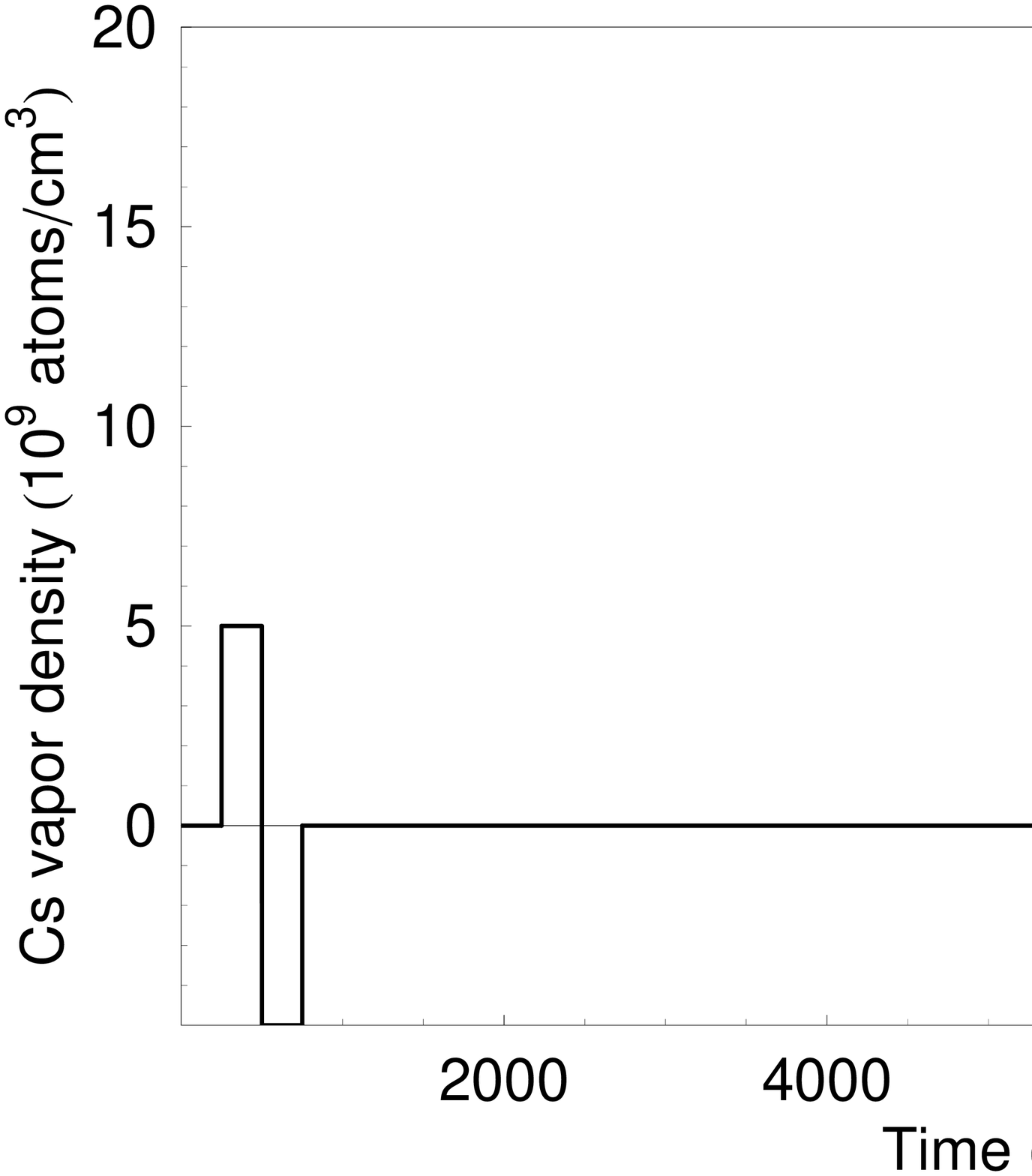}
\caption{Cesium density as a function of time (upper plots in each graph).  Applied electric field is shown in the lower plots in each graph.  Temperature of cell is $\approx 21^\circ$C.  In this experiment, the electric field polarity was switched at $|\vec{E}| = 5~{\rm kV/cm}$, and then the electric field was turned off for various intervals.  After this delay the field polarity was switched several times at $|\vec{E}| = 2~{\rm kV/cm}$.}
\label{Fig:Density_Efield_TimePlots_delay}
\end{figure}

Figure~\ref{Fig:FractionalDensityChange_vs_Efield_rev} shows the dependence of the maximum fractional decrease in Cs vapor density $(\Delta n/n_0)$ upon electric-field-polarity reversal as a function of the field magnitude for two experimental runs. In the first run, the experiment began at low electric field magnitude and the magnitude was subsequently increased in steps of 1~${\rm kV/cm}$, with 30 minute delays between changes in field magnitude.  In the second run, data were taken sequentially starting at 5~${\rm kV/cm}$ and subsequently decreasing the field magnitude.  We observed a ``hysteresis'' effect in the fractional density change:  the threshold for the appearance of the density drop is moved to a higher field magnitude for the run for which field magnitude was sequentially decreased.  This indicates that the effect is not merely related to the electric field magnitude, but also depends on the recent history of the cell.  A cell exposed to high electric field prior to a measurement exhibits a different response as compared to a cell that had no prior electric field applied.

To further elucidate this behavior, we performed the following experiment (the resulting data are shown in Fig.~\ref{Fig:Density_Efield_TimePlots_delay}).  We subjected a cell to a single polarity reversal with a field magnitude of $5~{\rm kV/cm}$ as shown in the lower part of the plots of Fig.~\ref{Fig:Density_Efield_TimePlots_delay}.  We then left the cell in zero field for a length of time $T$.  After this time, we subjected the cell to a series of field reversals of magnitude $2~{\rm kV/cm}$.  For delay times $T \lesssim 5000~{\rm s}$, a drop in density is observed only for the first polarity reversal.  After this there is no significant change in density.  However, for delay times $T \gtrsim 5000~{\rm s}$, the usual drop in density is observed for every polarity reversal.  This indicates a paraffin-coated cell can, for a time, be ``conditioned'' to allow field reversals without the usual drop in density, at least for electric field magnitudes in the neighborhood of the threshold field magnitude of $\sim 2~{\rm kV/cm}$.  This technique of pre-conditioning offers a possibility of conducting electric-field-related experiments with paraffin-coated cells while maintaining a relatively constant alkali vapor density.

\section{Experiments with Rubidium}\label{Sec:Rubidium}

The experiments discussed above provoke the questions of whether the observed effects are specific to cesium atoms or even whether the phenomenon is specific to the particular paraffin-coated vapor cells used.  Although there are established techniques to reliably produce high-quality paraffin-coated cells, there can be variations in the properties of paraffin-coated cells linked to details of their preparation and history.  For example, in some cells needle-like alkali metal crystals of lengths up to 3 mm have been observed to grow from the paraffin coating \cite{Bal07} (no such needles have been observed in the cells used in the present study).  Therefore it is of interest to study the application of electric fields to other paraffin-coated cells containing different atoms and using different experimental procedures to assess the generality of the phenomenon of electric-field-induced density changes.  Such experiments were carried out at RRI with a paraffin-coated rubidium vapor cell (containing an isotopically enriched sample of $^{85}$Rb) of similar size and shape as the cells used in the experiments with Cs at CSU-EB. Since the alkali atoms used, the electric field plate geometry employed, and the details of the experimental method all differ from the experiments with Cs carried out at CSU-EB, the two sets of experiments offer independent measurements of the effect.  Indeed an electric-field-induced change in the alkali vapor density, qualitatively similar to the CSU-EB results discussed above, was observed in the Rb cells.  Furthermore, the experimental approach at RRI enabled reliable measurement of the alkali atom ``burst'' occurring at the outset of an abrupt change of the electric field magnitude, allowing additional details of the effect to be studied.

\subsection{Experimental setup}

\begin{figure}
\center
\includegraphics[width=3.35 in]{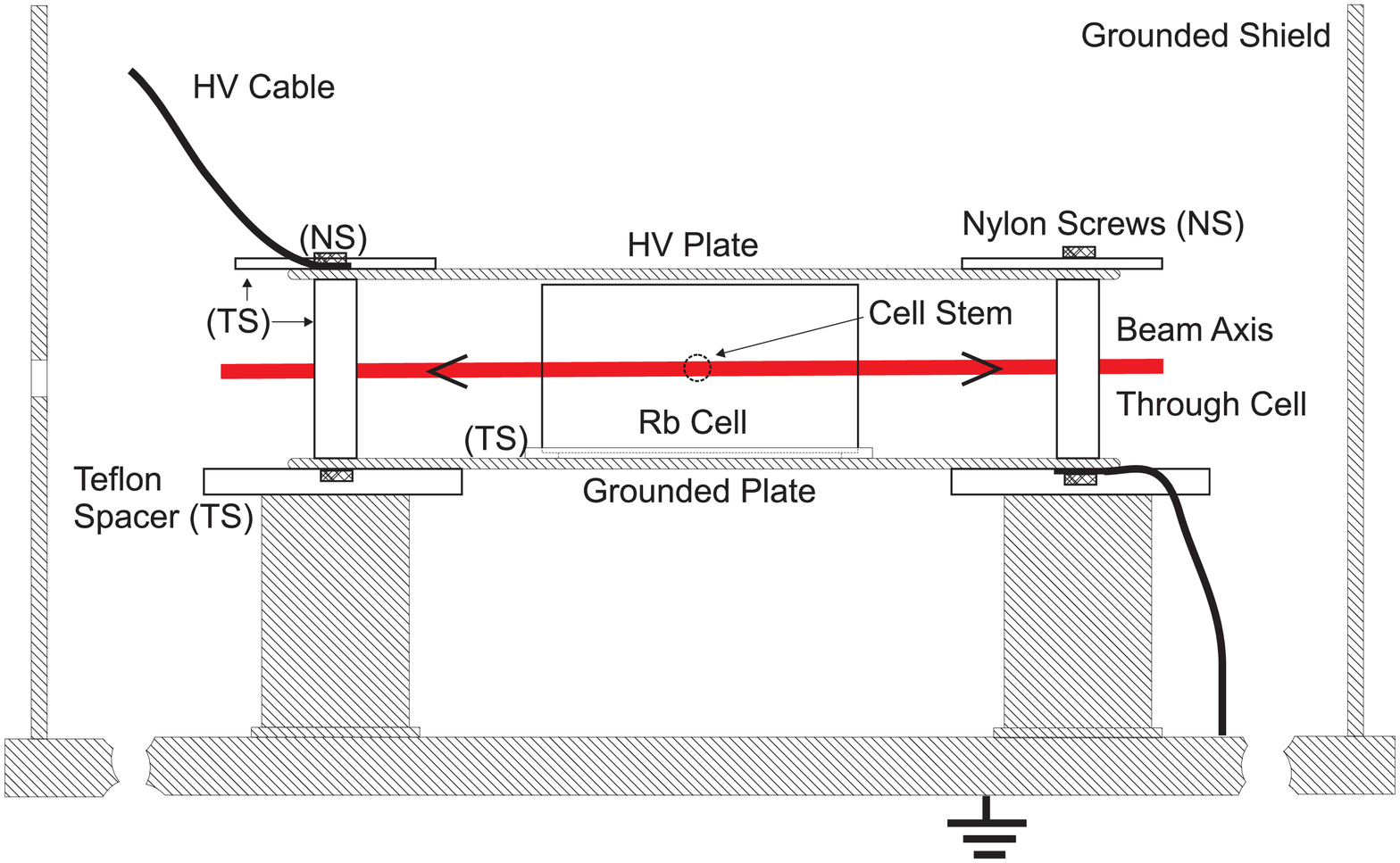}
\caption{Schematic diagram indicating the side view of the experimental arrangement for the Rb measurements. The insulating material is Teflon and all screws are Nylon. The beam path through the cell is indicated by the bold line. The stem of the cell is located at the far side of the cell in this view and is roughly parallel to the electric field plates.} \label{Fig:RRI_exp_setup}
\end{figure}

The experimental setup for the RRI measurements is shown in Fig.~\ref{Fig:RRI_exp_setup}.  The paraffin-coated Rb cell has a cylindrical geometry: the cell is 3.2~cm tall and 6.0~cm in diameter. The electric field across the cell is applied by two parallel plates of dimension 10.0~cm~$\times$~15.0~cm, held apart by 3.4~cm. The cell is placed symmetrically between the parallel plates, and teflon spacers prevent its direct contact with the electrodes (which is in contrast to the Cs experiments at CSU-EB where the electrodes were in direct contact with the cell). According to computer modeling of the electric field plate geometry, the electric field inhomogeneity across the volume occupied by the cell is expected to be less than 7\%. The Rb-containing stem of the cell is symmetrically situated between the plates. The lower plate is held at ground potential throughout the experiment while the upper plate takes on positive values of potential when the electric field is on.

An extended-cavity diode laser (ECDL) generating light at 780~nm near-resonant with the Rb D2 line was used to probe the vapor density of Rb in the cell. The grating of the ECDL was scanned with a triangular waveform at a rate of approximately 8~Hz over the Doppler width of the $F=3 \rightarrow F'$ component of D2 line of $^{85}$Rb, resulting in approximately 16 absorption cycles per second. Due to hysteresis of the ECDL grating's piezo element as a function of voltage, only one slope was sampled in the measurement for consistency. This restricts the time resolution of the experiment to $\approx$~130~ms (a factor of $\approx 50$ improvement over the time resolution of the Cs experiments at CSU-EB). To monitor the power and detuning of the laser light, the laser beam is split into two separate beams, a reference beam and a probe beam. The reference beam passes through an uncoated 7.5-cm long cell containing Rb in the natural isotopic abundance. The transmitted light intensity is recorded. The probe traverses the paraffin-coated Rb cell along the diameter of the circle in the cross-section of the cylinder. Both the probe and reference absorption measurements are done in double pass to enhance sensitivity to changes in vapor density.  The light power was sufficiently low for both reference and probe beams so that there was no saturation of the transitions, therefore the transmission measurements can be analyzed in the linear absorption regime.

Changes in absorption are measured by recording the probe beam transmission signal from which the percent absorption is determined.  Both reference and probe signals from the respective photodiodes are recorded simultaneously on two independent channels of a digital oscilloscope for a duration of 10 seconds (the reference signal is recorded primarily to correct for drifts in laser detuning).  The signal analyzed in these measurements is the time dependence of the fractional absorption of the probe normalized by the fractional absorption of the reference at the center of the Doppler-broadened $F=3 \rightarrow F'$ D2 resonance.  The Rb vapor density in the reference cell corresponds to the saturated vapor pressure (for these experiments the room temperature was $\approx 21^\circ$C corresponding to a saturated vapor density of $5 \times 10^9~{\rm atoms/cm^3}$), and thus from the ratio of these signals the vapor density in the paraffin-coated cell can be estimated assuming linear absorption.  The initial $^{85}$Rb vapor density in the paraffin-coated cell was found from this analysis to be $2 \times 10^9~{\rm atoms/cm^3}$, more than a factor of two smaller than the vapor density in the uncoated cell.

The time sequence of the measurement is as follows. The voltage required to produce the desired electric field between the plates is first pre-set, but not applied to the plates. The oscilloscope is then triggered to acquire the transmission signals. This gives the zero-field absorption spectra. Approximately one second into acquisition, the voltage to the electric field plates is turned on. Changes in absorption are recorded for the rest of the 10~s, while the electric field is maintained at a constant level across the cell. Eighty seconds to several minutes later the oscilloscope is triggered again. The first second of the acquisition occurs while the electric field is still at its previous level, to establish the steady state absorption level in the presence of the electric field. Approximately one second into acquisition, the voltage to the electric field plates is switched off and the transient response of the absorption is recorded for the remainder of the 10~s. Following this the next value of electric field is set and the measurements are repeated in the same manner with a delay of 90 minutes in between the measurements to allow for cell recovery (see Figs. \ref{Fig:FractionalDensityChange_vs_Efield_rev} and \ref{Fig:Density_Efield_TimePlots_delay} and surrounding discussion).

\subsection{Observations}

\begin{figure}
\center
\includegraphics[width=3.35 in]{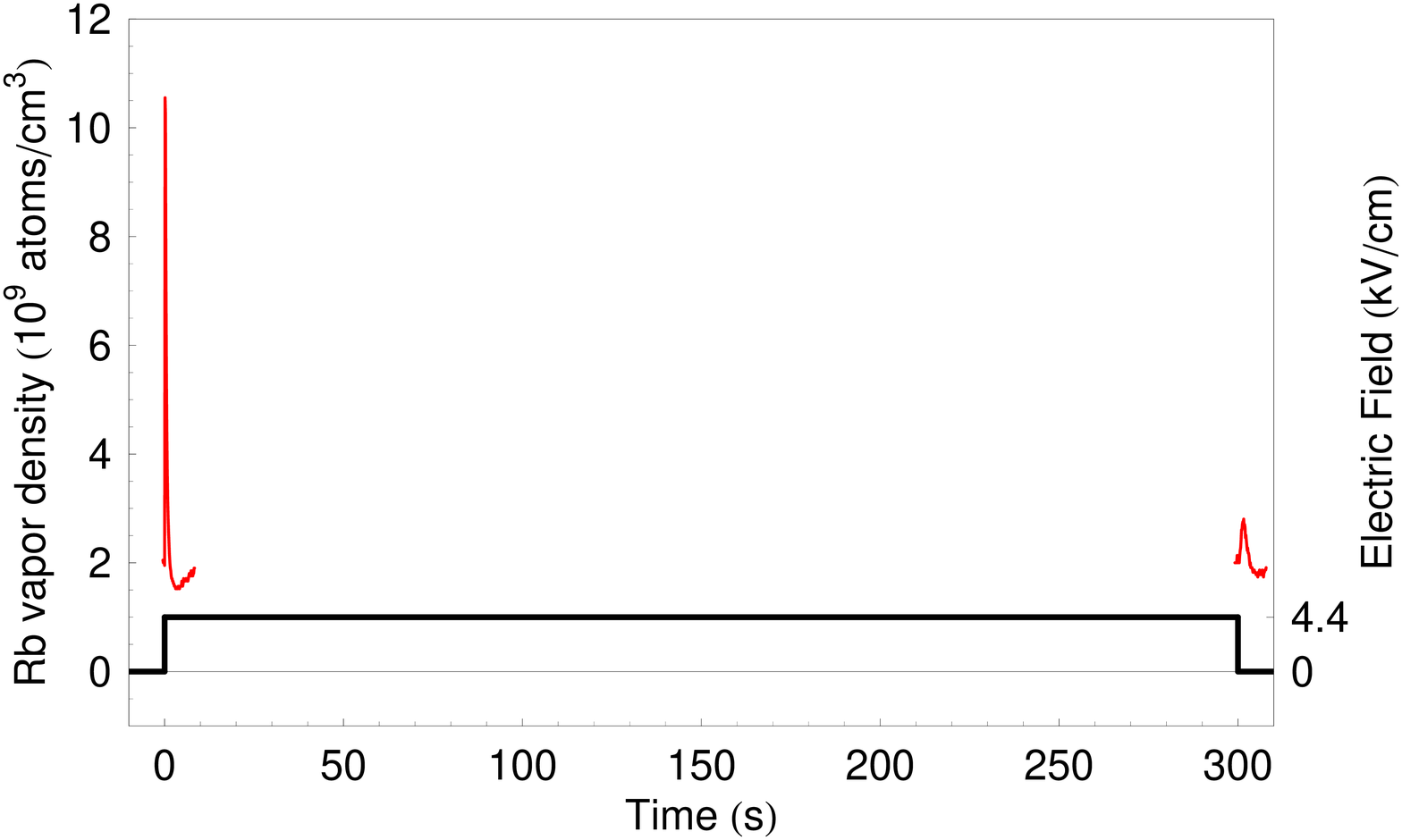}
\includegraphics[width=3.35 in]{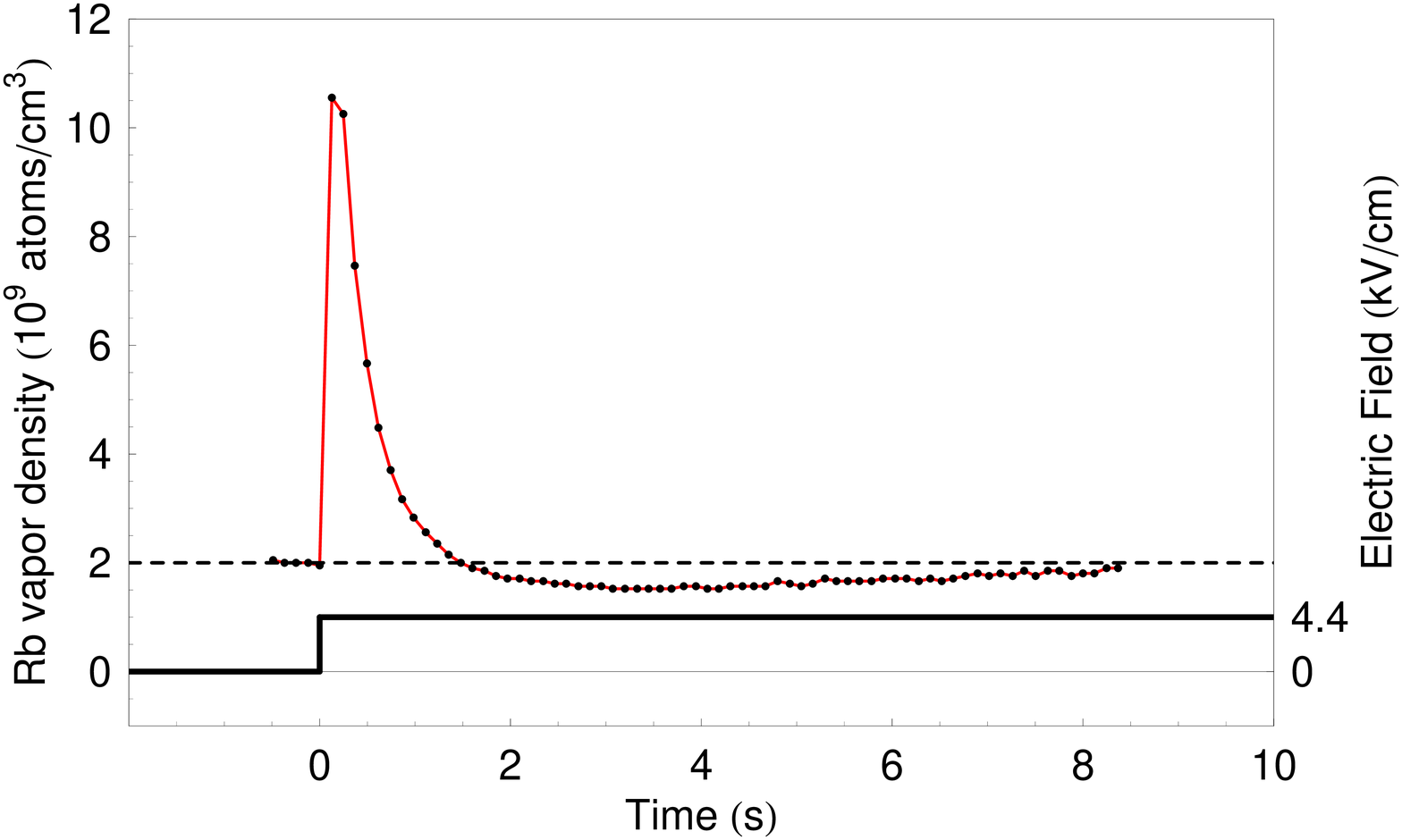}
\includegraphics[width=3.35 in]{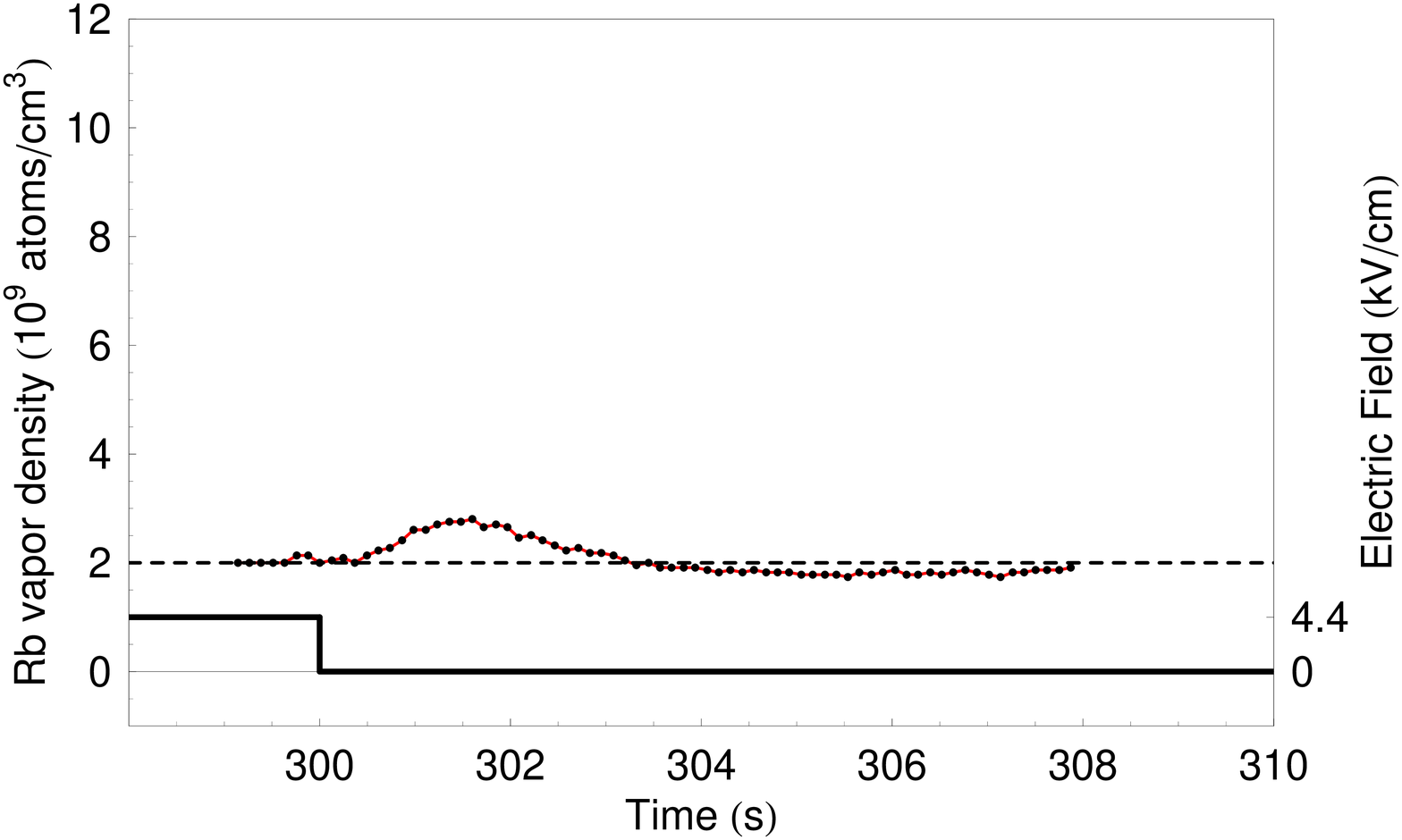}
\caption{Upper plot shows the time-dependent Rb vapor density in a paraffin-coated cell upon the sudden application of a 4.4~kV/cm electric field across it at time $t=0~{\rm s}$ and sudden switching off of the field at $t = 300~{\rm s}$.  The middle plot shows the data from the upper plot for the time-dependent Rb density during the turning on of the 4.4~kV/cm field with an expanded time scale, and the lower plot shows the vapor density during the turning off of the field with an expanded time scale.  The dashed horizontal lines represents the equilibrium vapor density in the cell at zero field.  The time interval between successive data points is $\sim 130~{\rm ms}$.  The rise time of the high voltage power supply for the electric field is $\sim 70~{\rm ms}$ and the fall time is $\sim 170~{\rm ms}$, comparable with the data acquisition rate.  The vapor density is calculated from the raw data by taking the natural log of the fractional light transmission signal and comparing with a calculation assuming linear absorption.  The absolute density level is accurate to within a factor of two, whereas the relative density level is accurate at the few percent level.} \label{Fig:RRI-transient-time-dependence}
\end{figure}

We first discuss a representative case of how the Rb vapor density in the paraffin-coated Rb vapor cell varies in time as the electric field applied to the paraffin-coated cell is turned on and off. Figure~\ref{Fig:RRI-transient-time-dependence} shows the time-dependent Rb vapor density when an electric field of 4.4~kV/cm is applied across the paraffin-coated Rb cell: the upper plot shows the full time sequence, the middle plot focuses on the vapor density as the high voltage is turned on, and the lower plot shows the vapor density when the high voltage is turned off.

In the middle plot of Fig.~\ref{Fig:RRI-transient-time-dependence}, the initial value is the steady-state value in the absence of the electric field. The sudden rise in vapor density (the ``burst'') follows immediately (time scale $\lesssim 100~{\rm ms}$) after the high voltage to the electric field plates is turned on. Following this, while the electric field is held at a constant value, the absorption declines (with a time constant of $\sim 1~{\rm s}$) falling below the zero-field value (the ``drop'') and then gradually returns to the zero-field value of absorption.  This behavior is qualitatively similar to the observations reported in Sec.~\ref{Sec:Cesium} for the paraffin-coated Cs cell when the polarity is switched, although the recovery time for the drop is considerably faster (5-10 times).  Also due to the superior time resolution of these experiments, the burst is clearly and consistently observed in the data.

Once the time-dependent absorption signal has settled again to a steady-state value (after between 80-120~s), the dynamics of the Rb vapor density when the electric field is switched off can be measured (Fig.~\ref{Fig:RRI-transient-time-dependence}, lower plot). A burst and drop in Rb density, measured through the absorption signal, are again observed when the electric field is switched off just as when it is switched on, although the amplitude and time constants are rather different. Upon switching the electric field off, the absorption increases gradually, peaking a little under two seconds after the switching of the field (note the fall time for the high voltage supply is $\sim 170~{\rm ms}$, much shorter than the observed time scale). Following this there is a gradual decline in the vapor density to below the initial equilibrium value, where a minimum is encountered. From here the absorption increases slowly again to its equilibrium value. Consistent with the earliest observations of such effects \cite{Kim01}, the burst and drop in vapor density is considerably larger for the field turning on as compared to the field turning off.

The long-term experimental trend is that the equilibrium number density is the same, with or without the electric field and this is fully recovered on a time scale significantly longer than the transients recorded above (several minutes).  A similar set of experiments performed with an uncoated alkali vapor cell resulted in no change in absorption whatsoever upon the turning on or off of the electric field. This again clearly indicates that the phenomenon relates to the paraffin coating.

\begin{figure}
\center
\includegraphics[width=3.35 in]{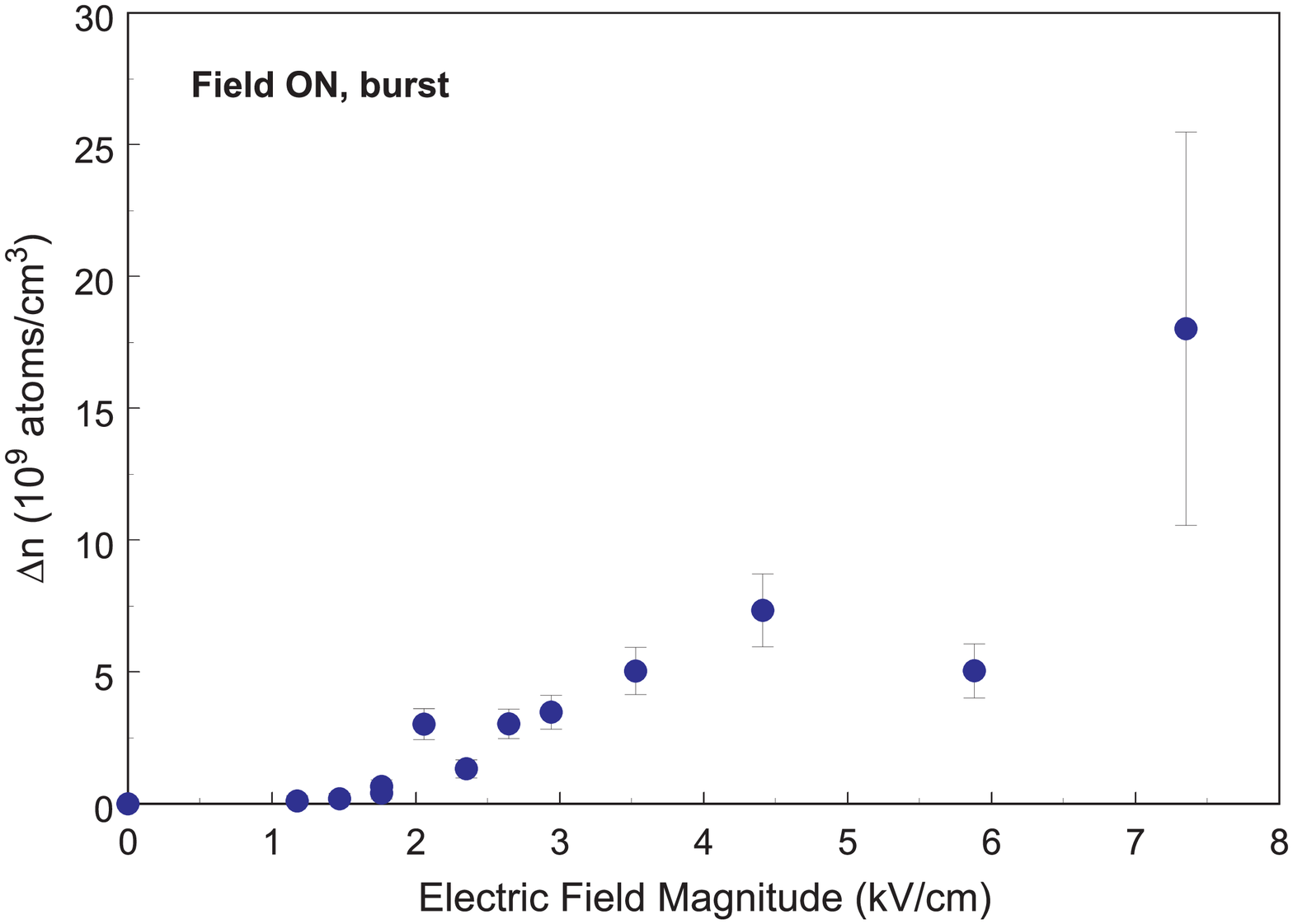}
\includegraphics[width=3.35 in]{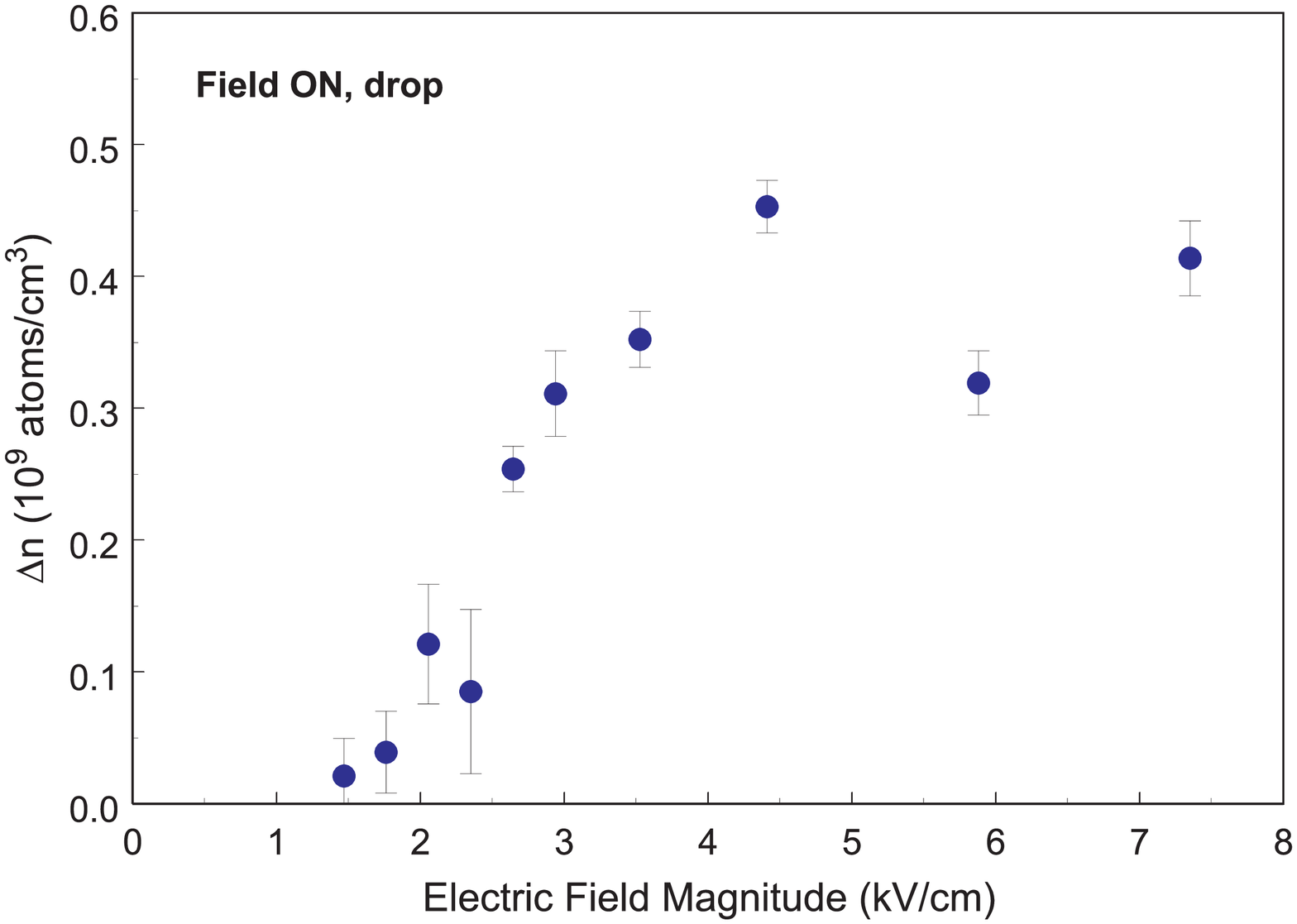}
\caption{Increase (upper plot) and decrease (lower plot) of the Rb vapor density in a paraffin-coated cell during the burst and drop of Rb vapor density, respectively, after the electric field is turned on as a function of applied electric field magnitude. Note that the vertical scales of the two plots are different by a factor of 20.  At approximately 1.6~kV/cm a threshold for the effect is observed. Note that two data points, one at $\approx$~2.1kV/cm and the other at $\approx$~6.0 kV/cm lie outside the trend, and these outliers did not reproduce in subsequent experiments.  These outliers are characteristic of the repeatability of the measurements.} \label{Fig:RRI-transient-ON}
\end{figure}

\begin{figure}
\center
\includegraphics[width=3.35 in]{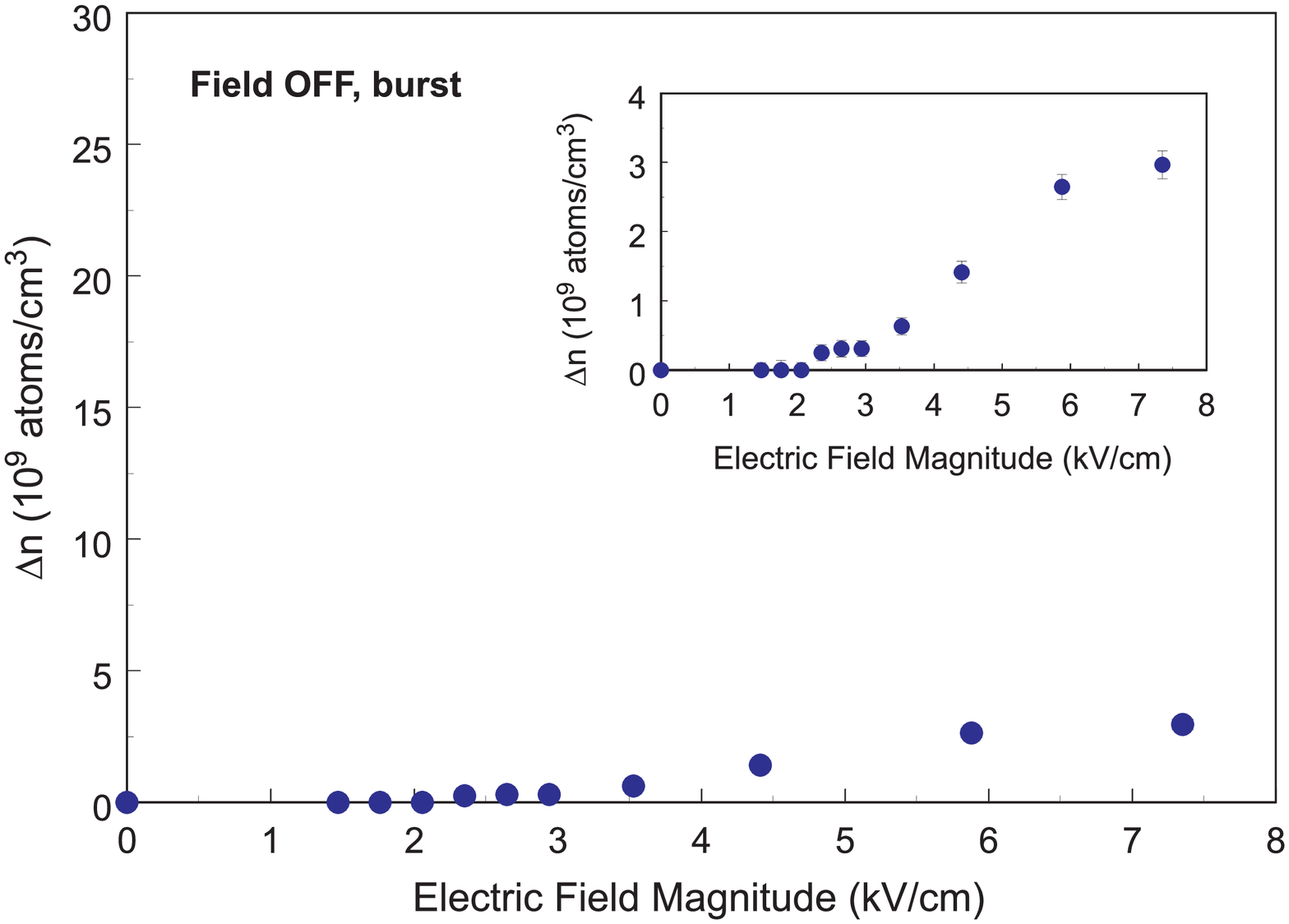}
\includegraphics[width=3.35 in]{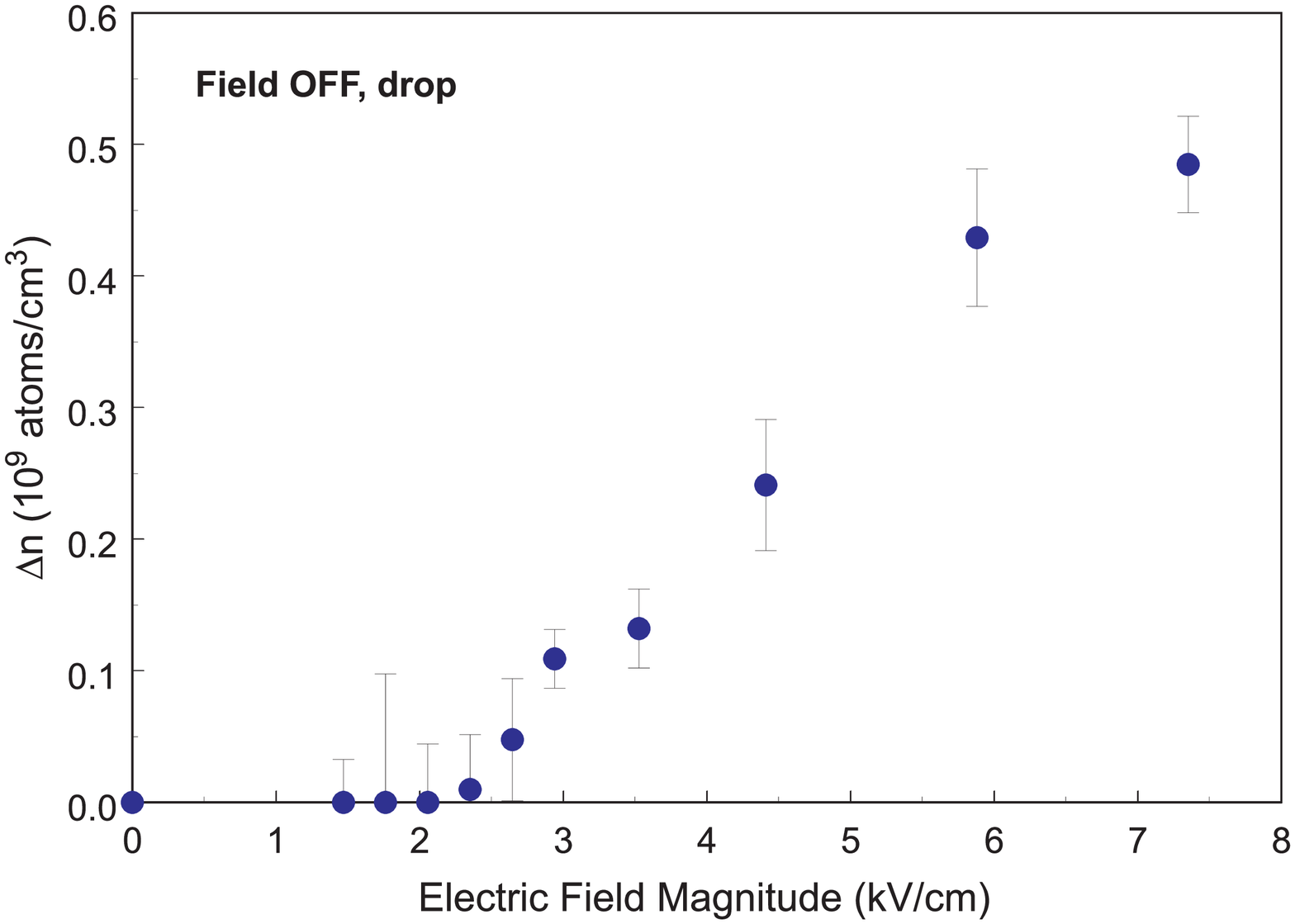}
\caption{Increase (upper plot) and decrease (lower plot) of the Rb vapor density in a paraffin-coated cell during the burst and drop of Rb vapor density, respectively, after the electric field is turned off as a function of applied electric field magnitude. The threshold for electric-field-induced Rb vapor density changes is at $\sim$~2.3~kV/cm, below which no effect is seen and above this value a steady increase is observed. Note that the burst amplitude for the field turning off is significantly smaller than that for the field turning on (Fig.~\ref{Fig:RRI-transient-ON}, upper plot), and to clearly observe the threshold an inset with expanding vertical scale is included in the upper plot.  The amplitudes of the drop in Rb density is within a factor of two of that observed for the turning on of the electric field throughout the data, and at high field magnitudes they are nearly equal.} \label{Fig:RRI-transient-OFF}
\end{figure}

The time-dependent absorption signals were recorded at several values of applied electric field across the cell. The range of electric field magnitudes in this experiment was from zero to 7.3~kV/cm. To track the systematic behavior of the Rb vapor density dynamics as a function of the applied electric field we plot the amplitude $\Delta n$ of the burst and drop in Rb vapor density for both the ``ON'' and ``OFF'' cycles. Below electric fields of 1.6~kV/cm, no statistically significant changes in Rb vapor density were observed. Above electric fields of 1.6~kV/cm, the peak absorption (and hence the peak Rb atomic density) during the burst tends to increase for the ON cycle (Fig.~\ref{Fig:RRI-transient-ON}, upper plot). Correspondingly, the magnitude of the drop increases as well (Fig.~\ref{Fig:RRI-transient-ON}, lower plot), although it should be noted that the drop $\Delta n$ is significantly smaller than the burst $\Delta n$ (please note that the vertical scales of the upper and lower plots of Fig.~\ref{Fig:RRI-transient-ON} differ by a factor of 45). While the burst $\Delta n$ appears to keep increasing as a function of increasing electric field above the threshold field magnitude, the drop $\Delta n$ appears to stabilize above electric fields of 5~kV/cm. In the OFF cycle (Fig.~\ref{Fig:RRI-transient-OFF}), the threshold for detectable changes in absorption shifts to higher values of electric fields as compared to the threshold values for the ON cycle. The error bars in the peak and minimum absorption signal plots (Figs.~\ref{Fig:RRI-transient-ON} and \ref{Fig:RRI-transient-OFF}) are statistical and are determined by shot-to-shot differences in the absorption amplitudes as well as fluctuation in the intra-shot absorption. Of these error estimates the electric-field-ON burst is the most imprecise as the response time for this effect is obviously shorter than the sampling time of the measurement (see upper plot of Fig.~\ref{Fig:RRI-transient-time-dependence}). This makes this measurement more susceptible to errors that are difficult to quantify. As all other response times are longer than the sampling rate, these have a more reliable error estimate.

Note that the amplitude of the vapor density drop at a field magnitude of $\approx 5~{\rm kV/cm}$ for Rb (lower plots of Figs.~\ref{Fig:RRI-transient-ON} and \ref{Fig:RRI-transient-OFF}) is $\sim 10^9~{\rm atoms/cm^3}$, whereas for Cs the observed vapor density drop at $\approx 5~{\rm kV/cm}$ is $\sim 10^{10}~{\rm atoms/cm^3}$ (Fig.~\ref{Fig:DensityChange_vs_Efield}).  This difference is near the difference in the initial vapor densities of the two species at room temperature.  However, note that changing the vapor pressure of the alkali atom in the cell through temporary heating appeared not to significantly alter the $\Delta n$ observed in the drop (Fig.~\ref{Fig:DensityChange_vs_Efield}).

The time constants for the various observed time-dependent absorption signals are independent of the value of the applied electric field magnitude once the respective thresholds are crossed, agreeing with the observations in Cs vapor cells discussed in Sec.~\ref{Sec:Cesium}.

Other experimental checks verified that the Rb vapor density changed also in the case when the electric field is slowly increased over a few seconds (the field magnitude is dialed up by hand) as opposed to the sudden switching of the electric field. In these measurements the threshold effect described above is also clearly evident at about the same magnitude of $\sim 2~{\rm kV/cm}$, however the amplitude $\Delta n$ of the burst and drop were affected by the switching time. Moving the cell around in the electric field region so that effects due to local inhomogeneities are brought out show no significant differences as compared to the reported results. In addition flipping the cell by 180 degrees with respect to the electric field direction also shows no deviations from the reported measurements.

\section{Effect of electric field on relaxation of atomic polarization}

An important question is whether the application of electric fields to paraffin-coated cells affects the relaxation of atomic polarization, since this is the primary reason to employ the paraffin-coated cells in experiments.  Another equally important question is whether or not there appears a systematic shift of ground-state Zeeman resonances associated with the observed phenomenon, since Zeeman resonance frequencies are the commonly measured quantities in fundamental symmetry tests and electromagnetic field measurements.  To investigate both the dependence of the relaxation rate and any Zeeman sublevel shifts associated with the electric fields, we employed the technique of nonlinear magneto-optical rotation (NMOR) using the apparatus described in Refs.~\cite{Bud98} and \cite{Bud00sens} at the University of California at Berkeley (UCB).

Yet a different cylindrical paraffin-coated cell, this one 2-cm tall and 6~cm in diameter, containing both Rb and Cs, was employed in these measurements.  The vapor density of Rb and Cs in the cell displayed time-dependent behavior similar to that reported in Secs.~\ref{Sec:Cesium} and \ref{Sec:Rubidium} upon turning on and off and switching the polarity of an applied electric field \cite{Kim01}.

\begin{figure}
\center
\includegraphics[width=3.35 in]{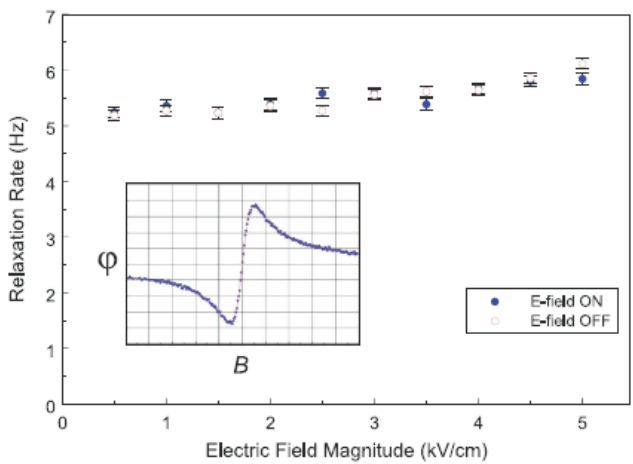}
\caption{Relaxation rate of $^{85}$Rb polarization in paraffin-coated cell with electric fields of various magnitudes on and off.  Relaxation rate is based on the width of nonlinear magneto-optical rotation resonances.  Data acquired after switching electric field and waiting for vapor density to reach equilibrium ($\approx 100~{\rm s}$).  Inset shows typical NMOR resonance data used to extract the relaxation rate $\gamma\ts{rel}$ from a fit of the data to Eq.~\eqref{Eq:NMORfit}.} \label{Fig:relaxationEfield}
\end{figure}

The cell was placed in the electric-field-plate apparatus shown in Fig.~\ref{Fig:HV electrodes}, and the electric-field-plate apparatus and cell were subsequently placed at the center of a four-layer ferromagnetic shield system with accompanying trim coils (described in Ref.~\cite{Bud98}).  The apparatus was at room temperature ($T \approx 20^\circ$C). Nonlinear magneto-optical resonances were observed using linearly polarized light from an external-cavity diode laser system, tuned to the high-frequency wing of the $^{85}$Rb $F=3 \rightarrow F'$ component of the D2 line.  The incident light power was $\approx 3.6~{\rm \mu W}$ and the laser beam diameter was $\approx 1~{\rm mm}$.  The direction of light propagation was along the axis of the electric field ($z$), and magnetic fields transverse to the light propagation direction were canceled with the trim coils to an accuracy of better than $0.1~{\mu}$G.  The magnetic field along $z$ was scanned, and the output polarization of the light was measured using modulation polarimetry (see, for example, \cite{Bud98} and \cite{Bud00sens}).  The resultant NMOR signals as a function of longitudinal magnetic field $B$ can be fit with the function (see, for example, \cite{NMOEreview})
\begin{align}
\varphi(B) = \alpha \frac{ \prn{ 2 g_F \mu_0 B / \gamma\ts{rel} }  }{ 1 + \prn{ 2 g_F \mu_0 B / \gamma\ts{rel} }^2 }~,
\label{Eq:NMORfit}
\end{align}
where $\alpha$ is the amplitude of the NMOR resonance (in general a function of vapor density, light power and detuning, etc.), $g_F$ is the Land\'{e} g-factor, $\mu_0$ is the Bohr magneton, and $\gamma\ts{rel}$ is the relaxation rate of atomic polarization in the cell.  An example of typical data and the corresponding fit is shown in the inset of Fig.~\ref{Fig:relaxationEfield}, where from the fit of the NMOR resonance $\gamma\ts{rel}$ was determined to be $\approx 5$~Hz.

\begin{figure}
\center
\includegraphics[width=3.35 in]{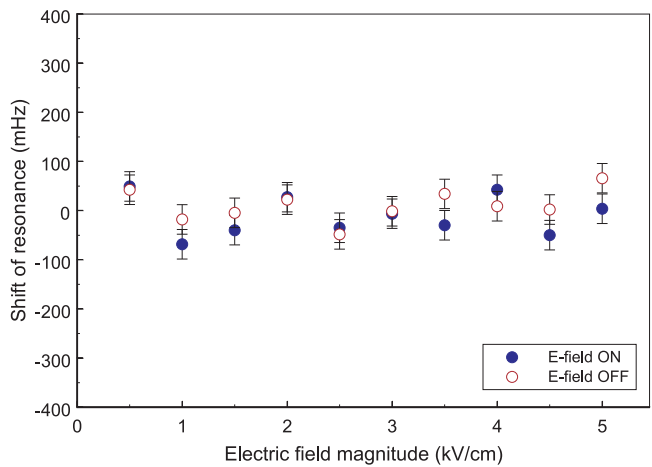}
\caption{Shift of the zero-field nonlinear magneto-optical rotation resonance for $^{85}$Rb $F=3 \rightarrow F'$ component of the D2 line with electric fields of various magnitudes applied to the paraffin-coated cell.  Data acquired after switching electric field and waiting for vapor density to reach equilibrium ($\approx 100~{\rm s}$).} \label{Fig:resonanceShiftEfield}
\end{figure}

Both the NMOR resonance width (Fig.~\ref{Fig:relaxationEfield}) and the value of the applied magnetic field $B$ along $z$ corresponding to zero field for the NMOR resonance (Fig.~\ref{Fig:resonanceShiftEfield}) were recorded as electric fields of various magnitudes were turned on and off (similar to the experimental procedure described in Sec.~\ref{Sec:Rubidium}).  The NMOR resonances were measured only after the vapor density had returned to near its equilibrium, zero-field value.

In general, at the level of sensitivity of these measurements, there is no statistically significant difference in the relaxation rates or shift of the zero-field resonance between measurements taken with the electric field on and those taken with the electric field off.  Thus one can conclude there is no catastrophic effect of the electric field on the properties of the paraffin coating essential for preservation of ground state atomic polarization, and it may still be possible to use such cells for precision measurements involving electric fields.  Indeed, the earlier electric dipole moment searches with paraffin-coated cells \cite{Ens62,Ens67,Eks71} and more recent experiments searching for the electric dipole moment of mercury employing cells with paraffin coating \cite{Rom01} support this finding. On the other hand, there does appear to be a statistically significant trend of increasing relaxation rate for measurements with the electric field both on and off (Fig.~\ref{Fig:relaxationEfield}) as the field magnitude is increased.  The electric field magnitude was increased sequentially in time during the measurement, so this could be a slight overall degradation of the cell's relaxation properties during the experiment, although there is insufficient data to conclude whether or not this degradation is electric-field related.  However, it should be noted that all the cited electric dipole moment experiments \cite{Ens62,Ens67,Eks71,Rom01} have observed degradation of the paraffin-coated cells during the experiments, which may be related to the effects discussed here.  Future experiments will address these issues in more depth.

\section{Interpretation}

\subsection{Overview of Observations}

The results of the experiments suggest that the observed electric-field-induced change in alkali vapor density in paraffin-coated cells is related to an electric-field-induced alteration of the properties of the paraffin coating.  Since the electric-field-induced changes in Rb and Cs vapor density are observed in paraffin-coated cells but not in uncoated cells, the effect must be related to the presence of the paraffin coating.  Furthermore, the time scales associated with the recovery of the cell to its initial character after application of an electric field (Fig.~\ref{Fig:Density_Efield_TimePlots_delay}) are too long to be associated with a purely gas-phase interaction --- there must be some relatively long-term modification of the paraffin.

A striking characteristic of both the burst and the drop in alkali vapor density induced by the electric field is the characteristic ``threshold'' for the effect (or, at least, a nonlinear dependence on electric field magnitude of the amplitude of the burst and drop in vapor density in the $\sim 0 - 2 {\rm kV/cm}$ range).  Figures~\ref{Fig:Density_Efield_TimePlots_20C}, \ref{Fig:DensityChange_vs_Efield}, \ref{Fig:FractionalDensityChange_vs_Efield_rev}, \ref{Fig:RRI-transient-ON}, and \ref{Fig:RRI-transient-OFF} demonstrate that significant changes in alkali vapor density only occur for field magnitudes above $1.6 - 2.2~{\rm kV/cm}$.  Note that the experiments demonstrate that the threshold is essentially independent of temperature (in the range $15^\circ$C - $25^\circ$C), alkali vapor density (in the range $7 \times 10^{9}$ - $3 \times 10^{10}~{\rm atoms/cm^3}$), and species of alkali atom (the threshold is similar for both Rb and Cs). The electric-field-magnitude threshold is evident for both the burst and drop in alkali vapor density (Figs.~\ref{Fig:RRI-transient-ON} and \ref{Fig:RRI-transient-OFF}).  The accumulated evidence indicates that the electric-field-induced change in alkali vapor density is caused by an alteration of the paraffin coating by the electric field, an alteration which has little to do with the density of alkali atoms in the cell.

Perhaps the effect is due to a physical restructuring of the arrangement of paraffin molecules in the coating by the electric field.  In the absence of an electric field, the paraffin molecules ($\rm{C_nH_{2n+2}}$, where $\rm{n} \sim 40-60$) are attracted to one another by van der Waals forces and form a tangled web on the glass surface.  Although rough estimates based on the surface tension and polarizability of paraffin \cite{CRC,Boy87} suggest that physical alteration of the paraffin coating is minimal over the studied range of electric field magnitudes, it may be possible that a small subset of weakly bound paraffin molecules begin to alter their configuration as the field magnitudes exceed the threshold value.

But why would a physical restructuring of the paraffin coating affect the vapor density in the cell?  Experiments using light to desorb alkali atoms from paraffin coatings \cite{Ale02,Gra05,Goz08,Kar08} have demonstrated that in paraffin-coated vapor cells a relatively large population of alkali atoms are adsorbed into the paraffin coating and are weakly bound to the paraffin molecules.  If the electric field physically restructures the coating, atoms previously trapped in the coating can be released, causing the burst in vapor density, and new binding sites for alkali atoms can be created, leading to the subsequent drop in vapor density.

In addition to the various lengths of the paraffin molecules, for which the surface tension and polarizibilities will be different, it is likely that there is considerable variation in binding energy of paraffin molecules due to their different locations and arrangements in the coating.  Thus it would seem plausible that once the threshold field magnitude is crossed, as the field magnitude is further increased a greater number of paraffin molecules will be re-oriented when the field is changed.  This offers an explanation for why the amplitude of the vapor density change increases as the field magnitude is increased past the threshold: larger electric field magnitudes lead to more re-oriented paraffin molecules, which in turn leads to more alkali atoms released from old binding sites and more new binding sites opening up.

In further support of this crude model, we note that the amplitude $\Delta n$ of the electric-field-induced drop in vapor density is independent of the initial vapor density (Fig.~\ref{Fig:DensityChange_vs_Efield}).  This is consistent with the appearance new binding sites in the paraffin coating, the number of which is unrelated to the equilibrium vapor density.

Lastly, the curious hysteresis behavior shown in Figs.~\ref{Fig:FractionalDensityChange_vs_Efield_rev} and \ref{Fig:Density_Efield_TimePlots_delay} is indicative of a physical restructuring of the paraffin coating that persists even once the electric field is set to zero, with a very slow relaxation back to a configuration similar to its initial configuration.  It has been observed \cite{Cha84} that paraffin can exhibit electret behavior, persistent polarization even in zero electric field.  The hysteresis effects observed could be connected to such persistent ``electret'' polarization.

This mechanism of restructuring paraffin with electric fields resulting in creation and passivation of binding sites for alkali atomic vapor offers a consistent mechanism for the effects seen with both the dipolar and the monopolar switching of the electric field. It is also consistent with the observation of the burst and drop when the electric field is switched off in the experiments with Rb. It is also important to note that different dynamic behavior is likely in paraffin with bipolar and monopolar switching. This could directly affect the number density of binding sites created, a fact that could explain the relative difference in the magnitude of the effect between the Cs and Rb experiments.

However, the model as presented is based on qualitative rather than quantitative features of the experimental observations, so further experimental and theoretical study is necessary to determine if this model is a viable explanation.

Further investigation might include experiments with different alkali atoms, other types of coatings, and possibly using light-induced atomic desorption (LIAD) \cite{Ale02,Gra05,Goz08,Kar08} in conjunction with electric fields to study the relationship between the effects.  Another important question left unanswered by the present set of experiments is the role, if any, of the rate of change of the electric field magnitude and direction.  The present experimental setup did not offer a convenient method of controlling the rate of change of the electric field $dE/dt$, but we hope to explore the dependence of the burst and drop in alkali density on $dE/dt$ as part of future work on this phenomenon.

\subsection{Model of vapor density dynamics}

In Ref. \cite{Ale02}, a phenomenological rate equation model describing the adsorption and desorption of alkali atoms from paraffin coatings was introduced to describe the change in alkali vapor density in paraffin-coated vapor cells exposed to light.  This relatively simple model has been useful in describing LIAD dynamics in subsequent work \cite{Gra05,Kar08}, and can provide a framework for understanding the underlying physics of atom/coating interactions.  The basic dynamics of the electric-field-induced alkali vapor density changes observed in the present experiments can also be described by this model, offering a glimpse at the essential features of an explanation of the phenomenon.

Based on accumulated evidence \cite{Ale02}, it is known that to describe changes in alkali vapor density $n(t)$ in paraffin coated cells, alkali atoms in three regions must be considered: the stem, the volume of the cell, and the paraffin coating.  We assume that the vapor density in the stem of the cell, $n_s$, is constant, and that there is an exchange rate between the stem and the volume of the cell described by a constant $\xi$ (which has units of ${\rm{cm^3/s}}$). It is known that the paraffin coating adsorbs and desorbs alkali atoms: the flux of atoms from the coating is $\gamma_d N_c(t)$, where $N_c(t)$ is the number of atoms in the coating, and the flux of atoms adsorbed into the coating is given by $\rho n(t) \bar{v} A / 4$, where $\rho$ is the probability of an atom being adsorbed into the coating, $\bar{v}$ is the average velocity of the atoms, and $A$ is the surface area of the paraffin coating. The probability of adsorption into the coating $\rho$ is assumed to be determined by the population of sites on the coating surface that are favorable for creating quasi-molecular bonds.  The atoms in the coating are irreversibly lost at a rate of $\Gamma$ via diffusion to the glass walls or possibly chemical reactions with impurities in the coating.

The rate equations describing the atoms in the cell with the electric field turned off are
\begin{align}
V\dbydt{}n(t) & =  -\rho \frac{\bar{v} A}{4}~n(t) + \xi\prn{n_s-n(t)} + \gamma_d N_c(t) \nonumber \\
\dbydt{}N_c(t) & = + \rho \frac{\bar{v} A}{4}~n(t) - \Gamma N_c(t) - \gamma_d N_c(t)~.
\label{Eq:RateEqns}
\end{align}

According to Eq.~(\ref{Eq:RateEqns}), the initial density in the cell $n_0 \equiv n(0)$ is given by
\begin{align}
n_0 & = \frac{\prn{\Gamma+\gamma_d}\xi}{(\rho \bar{v} A \Gamma)/4 +
\prn{\Gamma+\gamma_d}\xi}~ n_s~, \label{Eq:n0}
\end{align}
and the initial number of atoms in the coating is
\begin{align}
N_c(0) = \frac{\rho \bar{v} A}{4\prn{\Gamma+\gamma_d}}~ n_0 ~. \label{Eq:nc0}
\end{align}

Our present experiments suggest that the change in alkali vapor density may be caused by the electric field physically restructuring the paraffin coating.  Sites on the coating surface that previously bound atoms no longer bind the atoms, so the atoms are released causing the ``burst'' in density.  The restructuring causes new binding sites to appear, leading to the subsequent ``drop'' in vapor density.  The greater the restructuring of the coating, the greater the alteration of binding sites, leading to the dependence of the phenomenon on the electric field magnitude.  

We can model these effects using the rate equations \eqref{Eq:RateEqns} by introducing a temporary increase in both the adsorption probability $\rho$ and the desorption rate $\gamma_d$ when the electric field magnitude or direction is changed.  (Although it is entirely possible that the change of paraffin properties in not uniformly distributed over the coating, one can think of $\rho$ and $\gamma_d$ as average values for the entire coating.) We assign $\rho$ and $\gamma_d$ a time dependence when an electric field is changed suddenly at time $t=t_0$:
\begin{align}
\rho(t) = \rho_0 \prn{1 + \delta_a e^{-(t-t_0)/\tau_a}}~,
\label{Eq:rho-time-dep}
\end{align}
\begin{align}
\gamma_d(t) = \gamma_{d0} \prn{1 + \delta_d e^{-(t-t_0)/\tau_d}}~,
\label{Eq:gamma-d-time-dep}
\end{align}
where $\tau_a$ and $\tau_d$ are time constants characterizing the return of the parameters to their initial values, $\delta_a$ and $\delta_d$ are the amplitudes of the changes of the parameters, and $\rho_0$ and $\gamma_{d0}$ are their initial values in the absence of an electric field.  If the parameters $\rho$ and $\gamma_d$ did not relax back to their initial values, the vapor density $n(t)$ would not relax back to its initial value $n(0)$ as is observed in experiments.  The increase in $\gamma_d$ causes the burst in density and the increase in $\rho$ causes the drop in density.  Note that the functional form of the time dependence (for simplicity here chosen to be exponential) does not significantly affect the results of the model, since $\tau_a$ and $\tau_d$ turn out to be significantly shorter than other characteristic times associated with alkali vapor density dynamics in the cells.

\begin{figure}
\center
\includegraphics[width=3.35 in]{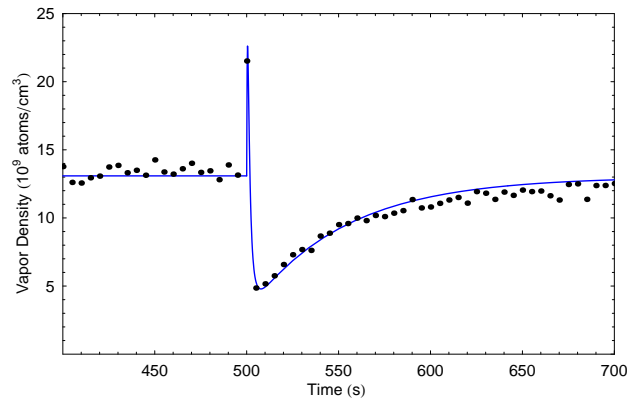}
\caption{Dots are experimental data for Cs vapor density in paraffin-coated cell upon switching of electric field polarity at time $t=500~{\rm s}$, shown previously in Fig.~\ref{Fig:Efield_reversal_4kV_Zoom}.  Solid line shows numerical solution of rate equations for $n(t)$ with the following parameters: $\rho_0 = 2 \times 10^{-6}$, $\delta_a = 100$, $\gamma_{d0} = 2 \times 10^{-4}~{\rm s^{-1}}$, $\delta_d = 1000$, $\tau_a = 2~{\rm s}$, $\tau_d = 0.5~{\rm s}$, $\xi=0.25~{\rm cm^3/s}$, $\Gamma = 5 \times 10^{-4}~{\rm s^{-1}}$, cell volume $V = 85~{\rm cm^3}$, coating surface area $A = 85~{\rm cm^2}$.  Parameters are in rough agreement with those from Refs.~\cite{Ale02,Gra05} where applicable.} \label{Fig:DataAndModel}
\end{figure}

\begin{figure}
\center
\includegraphics[width=3.35 in]{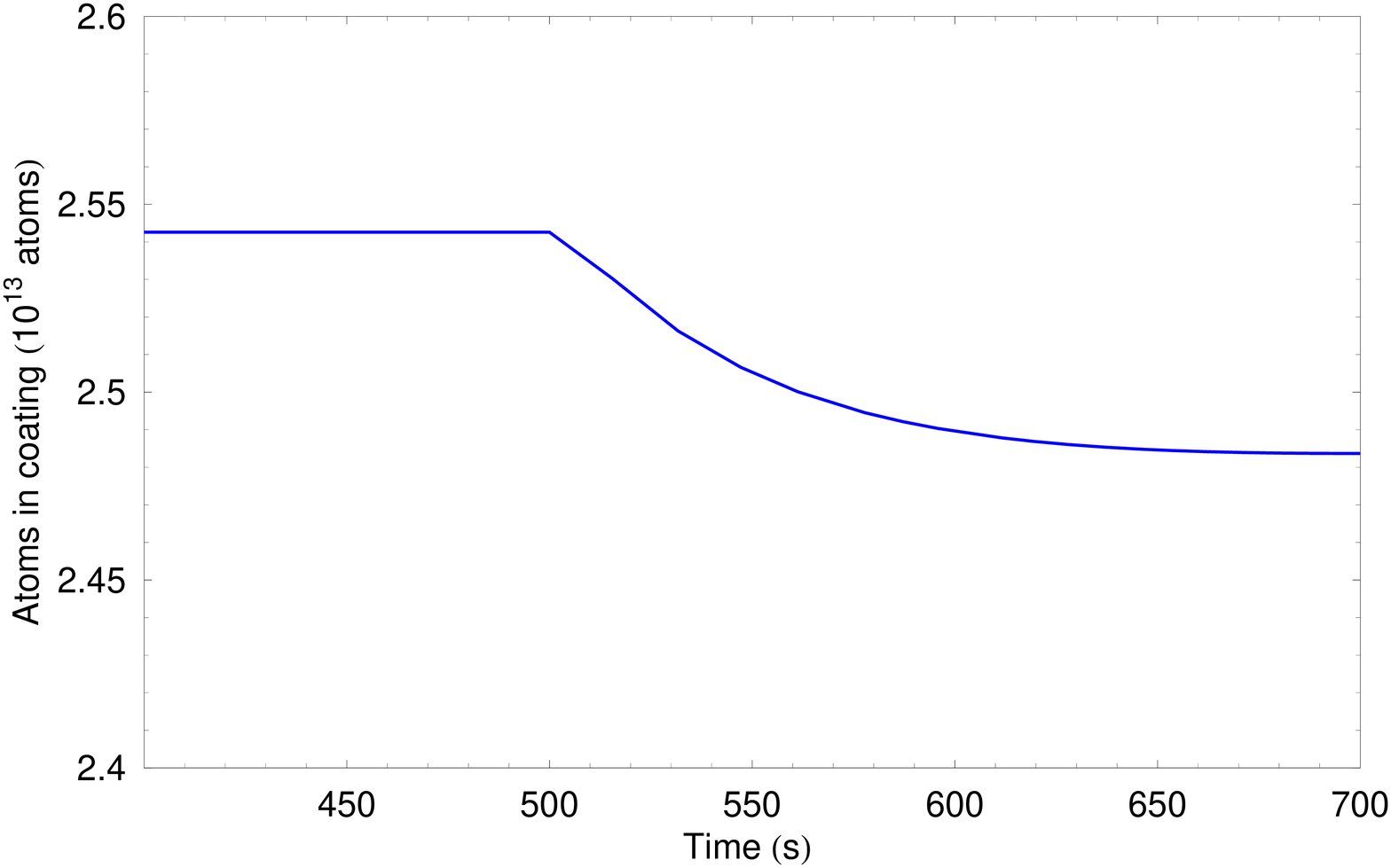}
\includegraphics[width=3.35 in]{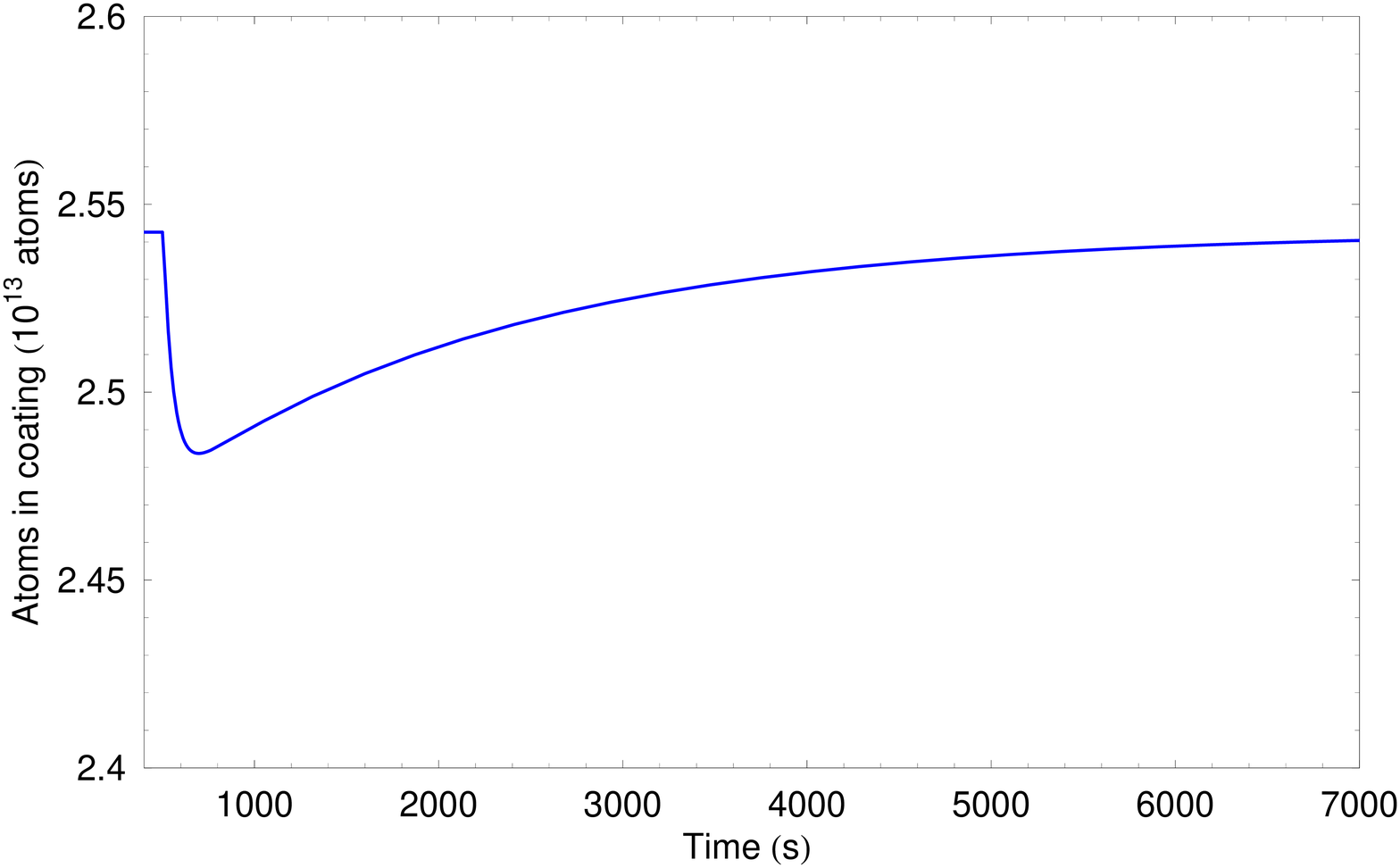}
\caption{Results of model calculation for parameters of Fig.~\ref{Fig:DataAndModel} for the number of atoms in the coating $N_c(t)$.  Upper plot has same time scale as Fig.~\ref{Fig:DataAndModel}, lower plot shows same results over a longer time scale to observe recovery of $N_c$ to its equilibrium value.  Note that the vertical scale does not begin at zero, and in fact there is less than a 5\% change in $N_c$ caused by the electric field.} \label{Fig:AtomsInCoating}
\end{figure}

Figure \ref{Fig:DataAndModel} compares the data shown previously in Fig. \ref{Fig:Efield_reversal_4kV_Zoom} to the results of a numerical solution of the rate equations \eqref{Eq:RateEqns} using the time dependent parameters from Eqs.~\eqref{Eq:rho-time-dep} and \eqref{Eq:gamma-d-time-dep}.  Parameter values, where possible, are chosen to be consistent with previous research on LIAD in paraffin-coated cells \cite{Ale02,Gra05,Goz08,Kar08}; values for parameters unique to the present research ($\tau_a$, $\tau_d$, $\delta_a$, and $\delta_d$) are chosen to optimize agreement with the data.  Figure~\ref{Fig:AtomsInCoating} shows the number of atoms in the coating after the switching of the electric field.

A few observations can be made based on the comparison of the rate equation model to the data.  The recovery of the alkali vapor density to its equilibrium value is primarily determined by the replenishing of the atoms in the cell by the metal drop in the stem, much as is the case for LIAD \cite{Kar08}, and thus the time scale of this recovery is unrelated to the electric-field-induced modification of the paraffin coating. (This is consistent with our observation that changing the electric field magnitude alters the amplitudes of the burst and drop in vapor density, but not the time scale of the burst and drop.)  As can be seen in Fig.~\ref{Fig:AtomsInCoating}, the time scale for recovery of the number of atoms in the coating is considerably longer than the time scale for recovery of the atomic vapor density, since the coating must refill through the relatively slow process of adsorption into the coating whereas the vapor density recovers on the considerably faster time scale set by the exchange of atoms between the cell volume and the stem.  Also of note is the fact that the phenomenon can be explained through rather brief (time constants $\sim 1~{\rm s}$) modification of the coating properties: vapor density changes persist for considerably longer periods of time because of the slow dynamics of re-equilibration of alkali atom population in the stem, coating, and volume of the cell.  (Nonetheless, there must also be a long-term change in the coating properties to explain dependence of the vapor density change on cell history as shown in Figs.~\ref{Fig:FractionalDensityChange_vs_Efield_rev} and \ref{Fig:Density_Efield_TimePlots_delay}).  Figure~\ref{Fig:AtomsInCoating} also indicates that although the effect is quite dramatic when observed through the vapor density, it can be explained with a relatively minor ($\lesssim$ 5\%) change in the number of atoms adsorbed into the coating.

Although the model is far from a complete explanation of the phenomenon, (for example, the model is not based on a first principles calculation, nor does it explain the longer-term modification of the coating properties demonstrated in Figs.~\ref{Fig:FractionalDensityChange_vs_Efield_rev} and \ref{Fig:Density_Efield_TimePlots_delay}), it does demonstrate that the observed phenomenon can be explained using a framework consistent with previous observations of LIAD in paraffin-coated cells \cite{Ale02,Gra05,Goz08,Kar08}.

\section{Conclusion}

We have carried out a variety of measurements of a heretofore unexplored electric-field-induced change of alkali vapor density in paraffin-coated cells.  Upon a change in the magnitude or polarity of an electric field (of sufficient magnitude, $\gtrsim 1.5~{\rm kV/cm}$) applied to a paraffin-coated cell, vapor density rapidly and dramatically increases (the ``burst''), then rapidly and dramatically decreases (the ``drop''), and then slowly recovers to its initial value.  The amplitude of the burst and drop in alkali vapor density were measured as a function of electric field magnitude, the cell temperature, equilibrium alkali vapor density, and previous exposure to electric fields.  Measurements were carried out in four different cells in three different laboratories, with qualitatively similar results.

A signature characteristic of the observed electric-field-induced change in alkali vapor density is an apparent threshold in electric field magnitude for the effect of between $\approx 1.6 - 2.2~{\rm kV/cm}$ which is independent of the alkali species, vapor density, and coating temperatures studied.  Another signature feature is a change in the effect after previous exposure of a cell to an electric field persisting for time scales of around an hour.

Because of their ability to preserve ground-state atomic polarization for exceptionally long times, the principle application of paraffin-coated cells is to precision atomic spectroscopy in magnetometers, clocks, and tests of fundamental symmetries.  The observations reported here have particular relevance to searches for parity- and time-reversal-violating permanent electric dipole moments carried out with paraffin-coated cells.  We confirmed that there are no significant changes in the relaxation properties of the coating, and no detectable shifts of Zeeman resonance frequencies.

\acknowledgments

We would like to thank B. P. Das and E. Krishnakumar for facilitating this work and Arun Roy and N. V. Madhusudana for helpful discussions.  We would also like to acknowledge the contribution of Morey Roscrow, Jr. to early parts of the experiment and the excellent technical assistance of Mohammad Ali and Alex Vaynberg in building parts of the apparatus.  This work has been supported by grant PHY-0652824 from the National Science Foundation and NSF/DST grant PHY-0425916 for U.S.-India cooperative research.

\end{document}